\pdfoutput=1
\documentclass[10pt,a4paper]{article}
\usepackage[T2A]{fontenc}
\usepackage[utf8]{inputenc}
\usepackage[english]{babel}
\usepackage{amssymb,graphicx}
\usepackage{amsmath,amsfonts}
\usepackage{amsthm}
\usepackage{framed}
\usepackage{fullpage}
\usepackage[makeroom]{cancel}
\usepackage[export]{adjustbox} 
\usepackage{feynmf}
\usepackage{microtype}
\usepackage{hyperref}

\newcommand{\sh}{\mathop{\mathrm{sh}}\nolimits}
\newcommand{\ch}{\mathop{\mathrm{ch}}\nolimits}
\renewcommand{\th}{\mathop{\mathrm{th}}\nolimits}

 \newcommand{\Tr}{\mathop{\mathrm{Tr}}\nolimits}

 \DeclareMathOperator{\sgn}{sgn}

\title{\textbf{$4d$ higgsed network calculus and elliptic DIM algebra }}
\author{Mohamed Ghoneim$^{1,2}$\thanks{mohamed.ghoneim@uni-bonn.de}, Can
  Koz\c{c}az$^{1,3}$\thanks{can.kozcaz@boun.edu.tr}, Kerem
  Kur\c{s}un$^{1}$ \thanks{keremkursun@hotmail.com}, Yegor Zenkevich$^{\text{4--9}}$\thanks{yegor.zenkevich@gmail.com}\\
{$^1$\small\textit{Department of Physics, Bo\u{g}azi\c{c}i University, Istanbul, Turkey}}\\
{$^2$\small\textit{Department of Physics and Astronomy, University of Bonn, Bonn, Germany}}\\
{$^3$\small\textit{Feza G\"{u}rsey Center for Physics and Mathematics, Bo\u{g}azi\c{c}i University, Istanbul, Turkey}}\\
  {$^4$\small\textit{SISSA, via
      Bonomea 265, 34136 Trieste, Italy,}}\\
  {$^5$\small\textit{INFN, Sezione di Trieste,}}\\
  {$^6$\small\textit{IGAP, via Beirut 2/1, 34151 Trieste, Italy,}}\\
  {$^7$\small\textit{ITEP, Bolshaya Cheremushkinskaya street 25, 117218
      Moscow, Russia,}}\\
  {$^8$\small\textit{ITMP MSU, Leninskie gory 1, 119991 Moscow, Russia,}}\\
  {$^9$\small\textit{MIPT, Institutskii pereulok 9, 141700, Dolgoprudny, Russia}}} 
\date{}

\begin{document}

\maketitle
\vspace{-60ex}
\begin{flushright}
   ITEP/TH-34/20\\
  MIPT/TH-19/20\\
\end{flushright}
\vspace{48ex}

\begin{abstract}
  Supersymmetric gauge theories of certain class possess a large hidden nonperturbative symmetry described by the Ding-Iohara-Miki (DIM) algebra which can be used to compute their partition functions and correlators very efficiently. We lift the DIM-algebraic approach developed to study holomorphic blocks of $3d$ linear quiver gauge theories one dimension higher. We employ an algebraic construction in which the underlying trigonometric DIM algebra is elliptically deformed, and an alternative geometric approach motivated by topological string theory. We demonstrate the equivalence of these two methods, and motivated by this, prove that elliptic DIM algebra is isomorphic to the direct sum of a trigonometric DIM algebra and an additional Heisenberg algebra.
\end{abstract}

\section{Introduction}
\label{sec:introduction}
String theory provides an effective framework for studying
supersymmetric gauge theories in various dimensions using geometric
and algebraic tools. These constructions are connected to each other
by string dualities. Depending on the properties of the gauge theories
we would like to study in detail one construction could shed more
light compared to the others. Moreover, these constructions utilize
geometric and algebraic approaches and intertwine them. The present
paper aims to study further the connection between the elliptic
Ding-Iohara-Miki (DIM) algebras~\cite{DI, Miki} and topological string
theory motivated approaches.
  
We can construct supersymmetric gauge theories using webs of branes in
type IIB string theory. These webs are trivalent graphs on a plane,
and such graphs are shown to be identical to the toric diagrams that
are used to engineer the gauge theory in type IIA/M-theory
compactifications~\cite{Leung:1997tw}. The (refined) topological
vertex~\cite{Iqbal:2007ii} is employed to calculate the topological
string free energy for the toric Calabi-Yau threefold which is
identical to the instanton free energy after the proper identification
of the gauge theory parameters with the K\"{a}hler classes of the
geometry. Although the topological vertex is constructed as a
geometric device to count maps, the quantum algebraic structure behind
it was discovered in~\cite{Awata:2011ce}. In particular, two different
versions of the refined topological vertex were shown to be matrix
elements of the same intertwining operator of Fock representations of
the DIM algebra.

The DIM algebra also plays the central role in the $q$-deformed
version of the well-known Alday-Gaiotto-Tachikawa (AGT) relation. The
connection between the DIM algebra and supersymmetric gauge theories
have been exploited in various directions. The fiber-base duality in
geometric engineering~\cite{Katz:1997eq} or its type IIB
interpretation as $S$-duality is mapped~\cite{Zenkevich:2014lca} to
the Miki automorphism~\cite{Miki},~\cite{Bourgine:2018fjy}. DIM also
provides a universal way to understand the
$qq$-characters~\cite{Nekrasov:2017gzb}, \cite{Kimura:2015rgi} using
quiver gauge theories in different dimensions~\cite{Mironov:2016yue},
\cite{Bourgine:2016vsq}. A network matrix model is introduced by
composing DIM intertwiners according to the data encoded in toric
diagrams, leading to $q$-deformed Dotsenko-Fateev integrals
corresponding to $q$-deformed vertex operator algebras. Recently, the
Higgsing construction has been employed to calculate the holomorphic
blocks of $3d$ ${\mathcal N}=2^{*}$
theories~\cite{Zenkevich:2017ylb}. Higgsing involves tuning certain
mass and Coulomb branch parameters of a ``parent'' $5d$ theory to
reach the point on the moduli space where the Higgs and the Coulomb
branches meet. The $3d$ theory is nothing but the worldvolume theory
of the vortices in the Higgs phase of the parent $5d$ theory
\cite{Dorey:2011pa,Hanany:2004ea,triality}.

Recently a new approach for direct computation of the $3d$ holomorphic
blocks using the vector representation of the DIM algebra was
proposed, the higgsed network calculus~\cite{Zenkevich:2018fzl}. In
this approach one engineers the $3d$ theories by suspending D3 branes
between NS5 branes, as opposed to building the parent $5d$ theories
from a web of 5-branes, without D3 branes. In the 5-brane web we
associate a Fock space to each one of the 5-branes with central
charges that match the $(p,q)$ charges of the corresponding brane. The
web encodes the brane charges in terms of the slopes in the associated
trivalent graph. However, when a D3 brane is connected to an NS5
brane, it does not bend the NS5 brane. Correspondingly a vector
representation does not have central charges and can be identified
with a D3 brane. The intertwining operator playing the role of the
D3-NS5 brane junction, is $\Phi(w):
\mathcal{F}^{(1,0),q,t^{-1}}_u\otimes \mathcal{V}^q_w \to
\mathcal{F}_{tu}^{(1,0),q,t^{-1}}$, where $\mathcal{V}$ stands for the
vector representation and $\mathcal{F}$ is for Fock representation.

In this work, we lift this construction one dimension higher by
studying the elliptic deformation of the DIM algebra proposed
in~\cite{Saito} as well as employing the~\emph{compactification} of
the geometry in the direction of the 5-brane,
following~\cite{Hollowood:2003cv}. In fact, it was argued
in~\cite{Iqbal:2015fvd,Nieri:2015dts} that the compactification of the toric diagram
and the elliptic deformation are equivalent extension of the AGT
relation to a relation between $6d$ uplift of the superconformal gauge
theory and the elliptic version of the Virasoro algebra. The
elliptically deformed Dotsenko-Fateev (DF) integrals were shown to
reproduce the $6d$ instanton generating functions that were computed
using the refined topological vertex. In this work, we also study the
relation between the compactification and elliptic DIM algebra.

According to~\cite{Saito}, an elliptic deformation of any vertex
operator given in free field representation can be obtained
algorithmically. This is done by doubling the number of free bosons
and then rescaling their modes by a factor containing the elliptic
deformation parameter. The algorithm produces manifestly elliptic
results for correlators of the deformed vertex operators. For example,
before the elliptic deformation the two point functions between the
$q$-$W$-algebra screening charges are expressed through the so-called
$q$-Pochhammer functions. Once the screening charges are elliptically
deformed following the aforementioned prescription, their two point
functions turn out to be given in terms of the \emph{elliptic} gamma
functions, the natural generalizations of the $q$-Pochhammer
functions. On the other hand, if we compactify along a single 5-brane
with a number of intertwining (as well as dual intertwining)
operators, we do not obtain a manifestly elliptic expression. However,
once the flavor symmetries are gauged by gluing multiple compactified
5-branes, the DF integrands become elliptic functions, and match what
we compute using the elliptic deformation of the vertex operators.

Having worked out the $4d$ lift of the higgsed network, we investigate
further the structure of the elliptic DIM algebras. We propose an
alternative but equivalent interpretation to the elliptic
deformation~\cite{Saito} in terms of thermo field doubles. We
construct an isomorphism between the elliptic DIM algebra and a direct
sum of trigonometric DIM algebra and a Heisenberg algebra. Both the
trigonometric and the elliptic DIM algebras are endowed with a Hopf
algebra structure. We \emph{conjecture} that the coproducts are
related after a nontrivial Drinfeld twist.

The paper is organized as follows. In sec. \ref{sec:EllDeform}, we
start with a brief review of the elliptic deformation introduced
in~\cite{Saito}, and compute the lift of the vacuum expectation value
for different configurations of intertwining operators which we will
use as building blocks to obtain the partition functions of $4d$
theories by gluing them to each other.  In
sec. \ref{sec:compactification}, we will compute matrix elements for
the same type of insertions of intertwining operators to compute their
traces. Section \ref{sec:ellipt-intertw} is reserved to check that the
elliptic deformation of the intertwining operators between a vector
and a Fock space representations are indeed the intertwining operators
of the elliptic DIM algebra. In sec. \ref{sec:elliptic-dim-algebra},
we reinterpret the elliptic deformation of the DIM algebra from
thermal quantum field theory perspective and show that the elliptic
DIM algebra is just the product of two simpler algebras, the
trigonometric DIM algebra and a Heisenberg algebra. We argue that the
trace in the usual DIM algebra is just vacuum expectation value in the
elliptic DIM. In the Appendices, we collect some definitions,
conventions and technical details that we skip in the main text.

\section{Elliptic Deformation of Vertex Operators}
\label{sec:EllDeform}

A one parameter elliptic deformation of the DIM algebra was first
introduced in~\cite{FHHSY} based on quasi-Hopf twist with the aim to
realize the elliptic version of the Macdonald operator
of~\cite{Ruijsenaars}. The elliptic kernel function associated with
this difference operator is proposed
in~\cite{KNS}. The precise connection between the
elliptic deformation of~\cite{FHHSY} and the elliptic kernel function
remains still unclear.

Another elliptic deformation of the DIM algebra was presented
in~\cite{Saito} to establish the aforementioned missing connection to
the elliptic kernel function. This deformation does not only depend on
an extra parameter related to the ellipticity but also requires
doubling of the Heisenberg algebra used in free field representation
of the algebra. In~\cite{Iqbal:2015fvd,Nieri:2015dts}, this elliptic form of DIM algebra
was shown to be the relevant deformation in lifting the AGT conjecture
to $6d$ theory on the gauge theory side, and elliptic conformal blocks
on the CFT side. Moreover, at the special points on the moduli space
of the $6d$ theory the instanton partition function coincides with the
vortex partition function supported on co-dimension two subspaces of
spacetime~\cite{Nieri:2015dts}. The reduction of the instanton partition
function to the vortex partition functions was previously observed for
$5d$ bulk theory in~\cite{triality}, and ealier for $4d$ theories
in~\cite{Dorey:2011pa}. Recently, a direct approach to obtain the vortex
partition function using the intertwining operators of vector
representations of the DIM algebra has been proposed. Motivated by the
results of~\cite{Iqbal:2015fvd,Nieri:2015dts}, we propose an elliptic deformation of the
vector intertwiner (higgsed vertex) according to the elliptic
recipe of~\cite{Saito}.

An algorithmic approach to elliptic deformation of a vertex operator
with a free field realization was introduced in \cite{Saito}, and we
briefly review it here for completeness. Suppose that we have a vertex
operator of the form
\begin{equation}
X(z)=\exp\left( \sum_{n\geq 1} X_{-n}^{-} a_{-n} z^{n}\right)\exp\left(\sum_{n\geq 1}X_{n}^{+} a_n z^{-n}\right),\label{eq:85}
\end{equation}
where $X^{\pm}_{\pm n}$ are generically complex numbers, and the
operators $\{a_n\}_{n\in\mathbb{Z}\backslash\{0\}}$ satisfy the
$q$-deformed version of the Heisenberg algebra,
\begin{equation}
[a_m,a_n]=m\frac{1-q^{|m|}}{1-t^{|m|}}\delta_{m+n,0}.\label{eq:84}
\end{equation}

The elliptic deformation involves two steps: in the first one, we
deform the Heisenberg algebra~\eqref{eq:84} with an additional factor
depending on the elliptic parameter $p$ and introduce an additional
deformed Heisenberg algebra:
\begin{align}
[a_m,a_n]=m(1-p^{|m|})\frac{1-q^{|m|}}{1-t^{|m|}}\delta_{m+n,0}, \qquad [b_m,b_n]=m\frac{1-p^{|m|}}{(qt^{-1}p)^{|m|}}\frac{1-q^{|m|}}{1-t^{|m|}}\delta_{m+n,0}.
\end{align}
The new Heisenberg algebra commutes with the existing one,
$[a_m,b_n]=0$.

The second setup consists of writing down two auxiliary vertex
operator in terms of the deformed modes $\{a_n\}$ and $\{b_n\}$,
\begin{align}
X_{b}(p;z)&\stackrel{\mathrm{def}}{=} \exp\left(-\sum_{n\geq 1} \frac{p^n}{1-p^n}X^{-}_{n}b_{-n}z^{-n}\right)\exp\left(-\sum_{n\geq 1}  \frac{p^n}{1-p^n} X^{+}_{-n} b_n z^n\right),\label{eq:86}\\
X_{a}(p;z)&\stackrel{\mathrm{def}}{=} \exp\left(\sum_{n\geq 1} \frac{1}{1-p^n}X^{-}_{-n}a_{-n}z^n\right)\exp\left( \sum_{n\geq 1}\frac{1}{1-p^n}X^{+}_n a_n z^{-n}\right ).\label{eq:87}
\end{align}
The elliptic deformation $X(p;z)$ of the operator $X(z)$ is defined as
a product of the auxiliary operators~\eqref{eq:86} and~\eqref{eq:87}:
\begin{align}
X^{(e)}(z) \stackrel{\mathrm{def}}{=} X_{a}(p;z)X_{b}(p;z).
\end{align}

The (trigonometric) intertwining operator $\Phi(w)$ associated with
D3-NS5 junction~\cite{Zenkevich:2018fzl} is given by
\begin{equation}
\Phi (w) = e^{- \epsilon_2 Q} w^{\frac{P}{\epsilon_1}} \exp \left( - \sum_{n\geq 1} \frac{w^n}{n}
    \frac{1-t^{-n}}{1-q^n} a_{-n}  \right) \exp \left( \sum_{n\geq 1} \frac{w^{-n}}{n}
    \frac{1-t^n}{1-q^{-n}} a_n \right).\label{eq:88}
\end{equation}
where $\epsilon_1 = \ln q$, $\epsilon_2 = - \ln t$ and $P$ and $Q$ are
the momentum and momentum shift operators respectively. We will use a
graphical representation of the intertwining operators to make the
structure of intertwiner networks more transparent. We also emphasize
that the line in the diagram can be viewed as branes: vector
representations (dashed lines) are D3 branes and Fock representations
(solid lines) are 5-branes. We thus draw:
\begin{equation}
  \Phi(w) =\quad  \includegraphics[valign=c]{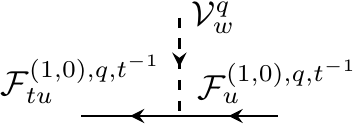}
\end{equation}

Note that the intertwining operator~\eqref{eq:88} is precisely in the
form required by the elliptic deformation algorithm reviewed above. We
thus apply the elliptic deformation to it as prescribed and thus
obtain the intertwiner $\Phi^{(e)}(w)$. It is straightforward to work
out the elliptic versions of a network with a pair of intertwining
operators:
\begin{equation}
  \label{s}
  \includegraphics[valign=c]{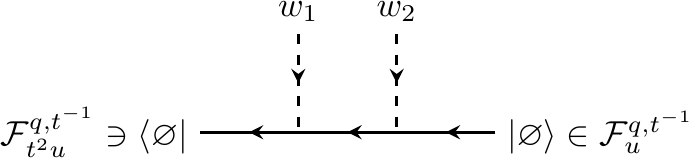} 
\end{equation}
We get
\begin{align}
\langle t^2 u,\varnothing|\Phi^{(e)}(w_1)\Phi^{(e)}(w_2)|u,\varnothing\rangle= w_1^{\log_q u+\beta}w_2^{\log_q u}  \frac{\Gamma_{p,q}(qw_2/w_1)}{\Gamma_{p,q}(qw_2/tw_1)},
\end{align}
where $\Gamma_{p,q}(z)$ is the elliptic gamma function. The
superscript $(e)$ indicates the elliptic version of the intertwining
operator. Note that in the limit when we turn off the elliptic
deformation by taking the limit $p\rightarrow0$, and using the fact
that the elliptic gamma function becomes inverse $q$-Pochhammer symbol
in this limit
\begin{align}
\Gamma_{p,q}(z)\rightarrow \frac{1}{(z;q)_{\infty}},\,\, \mbox{as}\,\,p\rightarrow 0,
\end{align} 
we can recover the original result
of~\cite{Zenkevich:2018fzl}. Furthermore, we can put this expression
in a nicer form using the reflection property of the elliptic gamma
function and a nice relation with the $\theta$-function:
\begin{align}
\langle t^2 u,\varnothing|\Phi^{(e)}(w_1)\Phi^{(e)}(w_2)|u,\varnothing\rangle= w_1^{\log_q u+\beta}w_2^{\log_qu} \, \frac{\theta_q(tw_1/w_2)}{\theta_q(w_1/w_2)}\frac{\Gamma_{p,q}(tw_1/w_2)}{\Gamma_{p,q}(w_1/w_2)}.
\end{align}
We can extend this result to $n$ insertions of intertwining operators
at points $w_1$, \ldots, $w_n$,
\begin{equation}
  \label{npoint1}
  \includegraphics[valign=c]{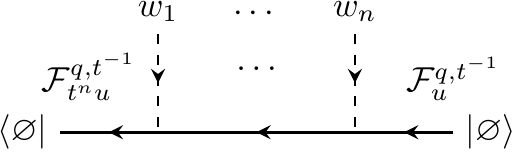}=
  A(\vec{w}) \left(
    \prod_{i=1}^n w_i^{\log_q u + \beta(i-1)} \right)  \prod_{i<j}
  \frac{\Gamma_{p,q}(tw_i/w_j)}{\Gamma_{p,q}(w_i/w_j)}
\end{equation}
where $A(\vec{w})$ remains invariant 
under the elliptic deformation,

\begin{equation}
  \label{sss}
  A(\vec{w})=\prod_{i<j}
  \left[ \left( \frac{w_i}{w_j} \right)^{\beta}  
  \frac{\theta_q\left( t
      \frac{w_i}{w_j}  \right)}{\theta_q\left( \frac{w_i}{w_j} 
    \right)} \right].
\end{equation}

The dual intertwining operator $\Phi^{*}(y)$ also has an explicit free
field representation
\begin{multline}
  \label{eq:93}
  \includegraphics[valign=c]{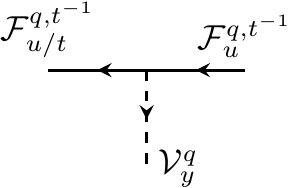}\quad =
  \Phi^{*}(y) =\\
  = e^{\epsilon_2 Q} y^{\beta - \frac{P}{\epsilon_1}} \exp \left[ \sum_{n\geq 1} \frac{y^n}{n}
    \left( \frac{t}{q} \right)^{\frac{n}{2}} \frac{1-t^{-n}}{1-q^n}
    a_{-n} \right] \exp \left[ - \sum_{n\geq 1} \frac{y^{-n}}{n}
    \left( \frac{t}{q} \right)^{\frac{n}{2}} \frac{1-t^n}{1-q^{-n}}
    a_n \right],
\end{multline}
and can be elliptically deformed by the above prescription. We denote
the resulting operator $\Phi^{(e)*}(y)$.

It is straightforward to compute the matrix element of the elliptic deformation of $n$ dual intertwining operators in a similar fashion,
\begin{equation}
\includegraphics[valign=c]{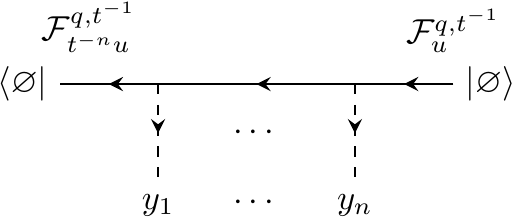}\quad =  \left(
  \prod_{i=1}^n y_i^{-\log_q u + \beta (i-n+1)}
\right) \prod_{i<j}
  \frac{\Gamma_{p,q}(ty_j/y_i)}{\Gamma_{p,q}(y_j/y_i)}.
\end{equation}

A combination of $n$ intertwining and $m$ dual intertwining operators
is another possible configuration that we will encounter in what
follows. We can order the intertwiners in two separate groups:
\begin{multline}
  \label{eq:94}
  \includegraphics[valign=c]{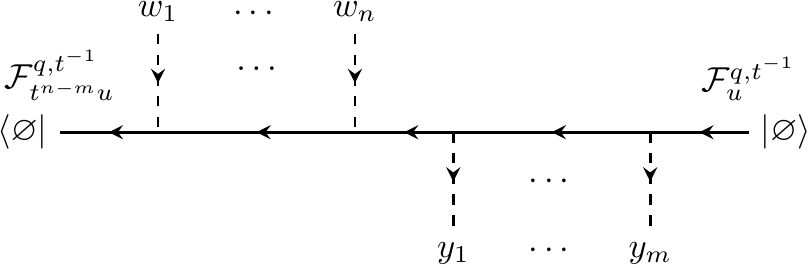} \quad = A(\vec{w}) B(\vec{w},\vec{y}) \left(
    \prod_{i=1}^n w_i^{\log_q u + \beta(i-2m-1)} \right)  \times\\
\times  \left(
  \prod_{i=1}^m y_i^{-\log_q u + \beta (i-2n+1)}
\right) \prod_{k<l}^n \frac{\Gamma_{p,q}\left(t \frac{w_k}{w_l} 
    \right)}{\Gamma_{p,q}\left(\frac{w_k}{w_l} 
    \right)}
  \prod_{i<j}^m\frac{\Gamma_{p,q}\left( t\frac{y_j}{y_i} 
    \right)}{\Gamma_{p,q}\left(t  \frac{y_j}{y_i}\right)}
  \prod_{a=1}^m \prod_{b=1}^n \frac{\Gamma_{p,q} \left(\sqrt{\frac{q}{t}}\frac{w_b}{y_a} \right)}{\Gamma_{p,q} \left(t\sqrt{\frac{q}{t}}\frac{w_b}{y_a} \right)},
\end{multline}
where an additional $q$-periodic prefactor reads 
\begin{equation}
  B(\vec{w},\vec{y}) =  \prod_{a=1}^m \prod_{b=1}^n \left[\left(
    \frac{y_a}{w_b} \right)^{\beta} \frac{\theta_q\left(  t\sqrt{\frac{q}{t}}
      \frac{y_a}{w_b} \right)}{\theta_q\left( \sqrt{\frac{q}{t}}
      \frac{y_a}{w_b} \right)}\right].
\end{equation}
Note that $A(\vec{w})$ and $B(\vec{w},\vec{y})$ remain invariant after
the elliptic deformation.

We also compute the commutators of the intertwining operators, which
will allow us to reorder the products like that in~\eqref{eq:94}:
\begin{enumerate}
\item Commutation of $\Phi^{(e)}(w_1)$ with $\Phi^{(e)}(w_2)$:
  \begin{equation}
  \includegraphics[valign=c]{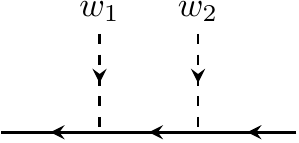} \quad = \left[
    \left(\frac{w_1}{w_2} \right)^{\beta} \frac{\theta_q \left( t \frac{w_1}{w_2} \right)}{\theta_q \left(
    \frac{w_1}{w_2} \right)} \right] R^{(e)} \left( \frac{w_1}{w_2} \right) \times \quad \includegraphics[valign=c]{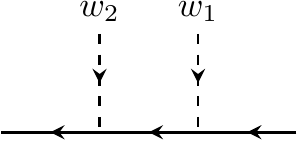}
\end{equation}
where we define the \emph{elliptic} $R$-matrix for vector
representations
\begin{equation}
  R^{(e)} \left( x \right) =  \frac{\Gamma_{p,q}\left(\frac{q}{t}
      x\right) \Gamma_{p,q}\left(t
      x\right)}{\Gamma_{p,q}\left(q
      x\right) \Gamma_{p,q}\left(
      x\right)}.
\end{equation}
Once again, the $q$-periodic prefactor remains invariant under the
elliptic deformation. The elliptic $R$-matrix $R^{(e)}$ tends to the
original trigonometric $R$-matrix~\cite{Zenkevich:2018fzl} in the limit
$p\rightarrow 0$.

\item Commutation of $\Phi^{*(e)}(y_1)$ with $\Phi^{*(e)}(y_2)$:
\begin{equation}
  \includegraphics[valign=c]{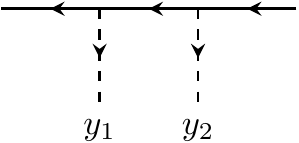} \quad =
  \left[\left( \frac{y_1}{y_2} \right)^{\beta}  \frac{\theta_q \left( \frac{y_2}{y_1} \right)}{\theta_q \left(
      t \frac{y_2}{y_1} \right)} \right] \frac{1}{R^{(e)} \left( \frac{y_1}{y_2} \right)} \times \quad \includegraphics[valign=c]{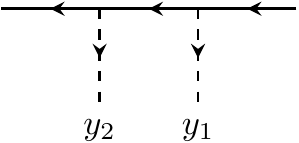}
\end{equation}

\item Commutation of $\Phi^{(e)}(w_1)$ with $\Phi^{*(e)}(y_2)$:
  \begin{equation}
  \includegraphics[valign=c]{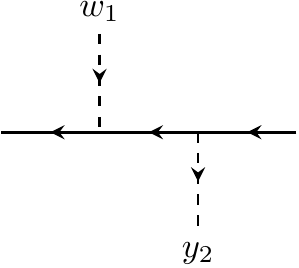} \quad =
  \left[\left( \frac{y_2}{w_1} \right)^{\beta}  \frac{\theta_q \left(
        t\sqrt{\frac{q}{t}}  \frac{y_2}{w_1} \right)}{\theta_q \left(
      \sqrt{\frac{q}{t}} \frac{y_2}{w_1} \right)} \right] \times \quad \includegraphics[valign=c]{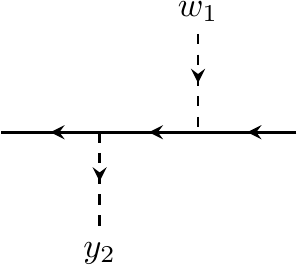}
\end{equation}
We find that this commutator is independent of the elliptic
deformation; hence, $\Phi^{(e)}(w_1)$ and $\Phi^{*(e)}(y_2)$
\emph{commute,} up to $q$-periodic factors. The periodic prefactors
will be irrelevant for us, since they factor out from all the
integrals in the partition functions we are going to consider.
\end{enumerate}

\section{Compactification}
\label{sec:compactification}
In the previous section, we have used the elliptic deformation of the
intertwining operators to compute the partition functions of the
theory lifted one dimension higher and compactified on a
circle. According to~\cite{Hollowood:2003cv}, this lift can be
obtained by wrapping the NS5 branes of the brane web around a
circle. The uplifted theory then lives on the D5 branes stretching
between the NS5 branes.

In this section, we will follow a similar approach for D3 branes in
the DIM language and show that it gives the same partition functions
as the elliptic deformation of sec.~\ref{sec:EllDeform}. As we have
mentioned earlier, there is a one-to-one correspondence between the
graphical representation of the intertwining operators and the brane
diagrams of type IIB string theory. Wrapping the 5-branes along a
circle requires the identification of the end points along the
horizontal direction. The compactification would algebraically
correspond to gluing the two ends, or equivalently computing the
trace,
\begin{multline}
  Z^{4d}(w_i;y_j)=\sum_{\mu}p^{|\mu|}b_{\mu}(q,t)\langle P_{\mu}| \Phi(w_1)\mathellipsis \Phi(w_n)\Phi^{*}(y_{1})\mathellipsis\Phi^{*}(y_{m})|P_{\mu}\rangle=\\
  =\mbox{Tr}_{{\mathcal F}}\left[ p^{L_{0}} \Phi(w_1)\mathellipsis
    \Phi(w_n)\Phi^{*}(y_{1})\mathellipsis\Phi^{*}(y_{m})\right]. \label{eq:89}
\end{multline}
where $b_{\mu}(q,t)$ are inverse norms of Macdonald polynomials and
$\mathcal{F}$ is the horizontal Fock representation. Notice that in
this section all the intertwiners are original \emph{undeformed}
operators~\eqref{eq:88},~\eqref{eq:93} and the ellipticity arises from
taking the trace of them.

The trace can be computed either using its cyclic property or by
explicitly evaluating the matrix elements in the first line of
Eq.~\eqref{eq:89}. Both computations lead to the same answer. For
completeness, let us write out the matrix elements. So far, we have
only computed vacuum-to-vacuum correlators, and now we need to find
the general matrix elements. Arbitrary matrix elements with $n$
insertions of the intertwining operators~\eqref{eq:88} can be computed
if one uses the Macdonald basis of states in the Fock space:
\begin{multline}
  \langle P_{\mu}| \Phi(w_1)\mathellipsis \Phi(w_n)|P_{\nu}\rangle
  \propto \frac{1}{b_{\mu}(q,t)}\prod_{ i<j}\frac{(q
    w_j/tw_i;q)_{\infty}}{(q w_j/w_i;q)_{\infty}}\times \\
  \times\sum_{\eta}Q_{\mu/\eta}(t^{-1}w_1,\mathellipsis,t^{-1}w_n;q,t)\,i_{w_{1}^{-1}}\mathellipsis
  i_{w_{n}^{-1}}P_{\nu/\eta}(qw_{1}^{-1},\mathellipsis,qw_{n}^{-1};q,t),
\end{multline}
where the proportionality sign appears because we have ignored the
zero modes associated with the operators. The involution $i_w$ is
defined by its action on the power sums, $i_wp_n(w)=-p_n(w)$. This
expression can be obtained by using the Cauchy identities and the skew
property of Macdonald polynomials. Likewise, we can compute the matrix
element of $m$ insertions of the dual intertwining operators
$\Phi^{*}$:
\begin{multline}
  \langle P_{\mu}|\Phi^{*}(y_{1})\mathellipsis \Phi^{*}(y_{m})|P_{\nu}\rangle\propto \frac{1}{b_{\mu}(q,t)}\prod_{i<j}\frac{(y_{j}/y_{i};q)_{\infty}}{(ty_{j}/y_{i};q)_{\infty}}\times\\
  \times \sum_{\eta}i_{y_{1}}\mathellipsis
  i_{y_{m}}Q_{\mu/\eta}(t^{-1}v^{-1}y_{1},\mathellipsis,t^{-1}v^{-1}y_{m};q,t)P_{\nu/\eta}(q
  v^{-1}y_{1}^{-1},\mathellipsis qv^{-1}y_{m}^{-1};q,t),
\end{multline}
where we used $v=q^{1/2}t^{-1/2}$ to simplify the notation. Having
found the arbitrary matrix elements with any number of insertions of
the intertwining operators and their duals, we can glue two such
pieces together to obtain the general \emph{strip} matrix elements:
\begin{multline}
  \langle P_{\mu}| \Phi(w_1)\mathellipsis \Phi(w_n)
  \Phi^{*}(y_{1})\mathellipsis\Phi^{*}(y_{m})|P_{\nu}\rangle\propto \\
  \propto \sum_{\eta}b_{\eta}(q,t) \langle P_{\mu}|
  \Phi(w_1)\mathellipsis \Phi(w_n)|P_{\eta}\rangle\langle
  P_{\eta}|\Phi^{*}(y_{1})\mathellipsis\Phi^{*}(y_{m})|P_{\nu}\rangle.
\end{multline}
Using the Cauchy identities we get
\begin{multline}
  \langle P_{\mu}| \Phi(w_1)\mathellipsis
  \Phi(w_n)\Phi^{*}(y_{1})\mathellipsis\Phi^{*}(y_{m})|P_{\nu}\rangle \propto\\
  \propto \frac{1}{b_{\mu}(q,t)}\prod_{1\leq i<j\leq
    n}\frac{(qw_{j}/tw_{i};q)_{\infty}}{(qw_{j}/w_{i};q)_{\infty}}\prod_{1\leq
    i <j\leq
    m}\frac{(y_{j}/y_{i};q)_{\infty}}{(ty_{j}/y_{i};q)_{\infty}}\prod_{i=1}^{n}\prod_{j=1}^{m}\frac{(t\sqrt{\frac{q}{t}}\frac{y_{j}}{w_{i}};q)_{\infty}}{(\sqrt{\frac{q}{t}}\frac{y_{j}}{w_{i}};q)_{\infty}}\times
  \\
  \times i_{y_{1}}\mathellipsis i_{y_{m}} i_{w^{-1}_{1}}\mathellipsis
  i_{w^{-1}_{n}}\sum_{\eta}Q_{\mu/\eta}(t^{-1}w_{1},\mathellipsis,t^{-1}w_{n},t^{-1}v^{-1}y_{1},\mathellipsis,t^{-1}v^{-1}y_{m};q,t)
  \times\\
  \times
  P_{\nu/\eta}(qw_{1}^{-1},\mathellipsis,qw_{n}^{-1},qv^{-1}y_{1}^{-1},\mathellipsis,qv^{-1}y_{m}^{-1};q,t).\label{eq:90}
\end{multline}

Using the matrix elements~\eqref{eq:90}, we can compute the
traces~\eqref{eq:89}. For example, the trace for the last strip with
$n$ insertions of the intertwining and $m$ dual operators is given by
\begin{multline}
  Z^{4d}(w_i;y_j)=\mbox{tr}_{{\mathcal F}}\left[ p^{L_{0}}
    \Phi(w_1)\mathellipsis
    \Phi(w_n)\Phi^{*}(y_{1})\mathellipsis\Phi^{*}(y_{m})\right] = \\
  =\frac{1}{(p;p)_{\infty}} \left [\frac{(pqt^{-1};p,q)_{\infty}}{(q;p,q)_{\infty}} \right ]^{n}\left [ \frac{(p;p,q)_{\infty}}{(pt;p,q)}\right]^{m} \prod_{1\leq i<j\leq n}\frac{(qw_{j}/tw_{i};p,q)_{\infty}}{(qw_{j}/w_{i};p,q)_{\infty}}\frac{(pqw_{i}/tw_{j};p,q)_{\infty}}{(pqw_{i}/w_{j};p,q)_{\infty}}\times\\
  \times\prod_{1\leq i<j\leq
    m}\frac{(y_j/y_i;p,q)_{\infty}}{(ty_j/y_i;p,q)_{\infty}}\frac{(py_i/y_j;p,q)_{\infty}}{(pty_i/y_j;p,q)_{\infty}}\prod_{i=1}^{n}\prod_{j=1}^{m}
  \frac{\Gamma_{p,q}(vy_j/w_i)}{\Gamma_{p,q}(vty_j/w_i)}\label{eq:91}
\end{multline}

The partition function~\eqref{eq:91} differs significantly from our
earlier result~\eqref{eq:94} where we employed the elliptic
deformation of the vertex operators. In contrast to Eq.~\eqref{eq:91}
not all factors in Eq.~\eqref{eq:94} can be packed into elliptic gamma
functions --- they are mostly expressed in terms of (double)
$pq-$Pochhammer symbols. It turns out that the situation here is
similar to the open topological string amplitudes appearing in
M-strings~\cite{Haghighat:2013gba}. In the refined case, open
amplitudes associated with the basic building blocks of the partition
function could not be written in terms of elliptic Jacobi
$\theta$-functions; however, if the pieces were glued together to
compute closed amplitudes the final answer could be recast entirely
using elliptic functions.
 
Following this observation we notice that if we take two compactified
strips~\eqref{eq:91} and glue them together \emph{on top} of each
other, the resulting expression \emph{can be written} in terms of
elliptic gamma functions:
 \begin{multline}
 Z^{4d}(w_i;y_j)Z^{4d}(y_j;z_k)\propto \prod_{1\leq i <j\leq m}\frac{\Gamma_{p,q}(qy_j/y_i)\Gamma_{p,q}(ty_j/y_i)}{\Gamma_{p,q}(y_j/y_i)\Gamma_{p,q}(qy_j/ty_i)} \prod_{i=1}^{n_{1}}\prod_{j=1}^{m} \frac{\Gamma_{p,q}(vy_j/w_i)}{\Gamma_{p,q}(vty_j/w_i)}\times\\
 \times \prod_{k=1}^{n_{2}}\prod_{j=1}^{m} \frac{\Gamma_{p,q}(vz_k/y_j)}{\Gamma_{p,q}(vtz_k/y_j)},\label{eq:92}
\end{multline}
where we have ignored all the $q$-periodic factors.

In fact, to complete the gluing procedure Eq.~\eqref{eq:92} has to be
integrated over the positions of the intermediate D3 branes $y_i$. The
resulting integral is the partition function of a $4d$ gauge theory
with $U(m)$ gauge group and $U(n_1)\times U(n_2)$ flavor groups. This
kind of integrals are familiar from the computations of the conformal
blocks using the $q$-deformed Dotsenko-Fateev Coloumb gas
representation. The factors appearing in the integrand originate from
Wick contractions of the screening currents with each other and with
the vertex operators. 


\section{Elliptic deformations of the intertwiners are intertwiners of
  the elliptic DIM algebra}
\label{sec:ellipt-intertw}
The elliptic deformation of the vector intertwiner $\Phi^{(e)}:
\mathcal{V}_q \otimes \mathcal{F}_{q,t^{-1}}(u) \to
\mathcal{F}_{q,t^{-1}}(t u)$ is given by (see
Appendix~\ref{sec:horiz-fock-repr} for the definitions of $a_n$,
$b_n$, $P$ and $Q$)
\begin{multline}
  \label{eq:95}
  \Phi^{(e)}(w) \stackrel{\mathrm{def}}{=} \Phi^{(e)} |w\rangle \otimes \ldots =\\
  = e^{-\epsilon_2 Q}
  w^{P/\epsilon_1} : \exp \left[ \sum_{n\neq 0} \frac{w^{-n}}{n}
    \frac{1-t^n}{(1- q^{-n})(1-p^{|n|})} a_n  + \sum_{n \neq 0}\frac{w^{n}}{n}
    \frac{(1-t^{-n})p^{|n|}}{(1- q^n)(1-p^{|n|})} b_n\right]:
\end{multline}
where $\epsilon_1 = \ln q$, $\epsilon_2 = - \ln t$ and we use the
basis $|w\rangle = \delta(x/w)$ in the representation
$\mathcal{V}_q$. The zero modes are the same as in the trigonometric
case, while the nonzero modes are obtained by the elliptic deformation
recipe~\cite{Saito} from the trigonometric
intertwiner~\cite{Zenkevich:2018fzl}.

It is straightforward to see that the operator~\eqref{eq:95} indeed
satisfies the intertwining relations
\begin{equation}
  g \Phi^{(e)} = \Phi^{(e)} \Delta(g)
\end{equation}
for $g$ any current $x^{\pm}(z)$ or $\psi^{\pm}(z)$.

There is also a dual intertwining operator $\Phi^{*(e)}:
\mathcal{F}_{1,t^{-1}}(u) \to \mathcal{F}_{1,t^{-1}}(u/t) \otimes
\mathcal{V}_q$ which reads
\begin{multline}
  \Phi^{*(e)}(w) \stackrel{\mathrm{def}}{=} ( \ldots \otimes \langle w|) \Phi^{*(e)}  =e^{\epsilon_2 Q}
  w^{\beta - P/\epsilon_1} \times\\
  \times  : \exp \left[ - \sum_{n\neq 0}
    \frac{w^{-n}}{n} \left( \frac{t}{q} \right)^{\frac{|n|}{2}}
    \frac{1-t^n}{(1- q^{-n})(1-p^{|n|})} a_n  - \sum_{n \neq
      0}\frac{w^{n}}{n} \left( \frac{q}{t} \right)^{\frac{|n|}{2}}
    \frac{(1-t^{-n})p^{|n|}}{(1- q^n)(1-p^{|n|})} b_n\right]:
\end{multline}
Note that the zero modes of the intertwiners $\Phi^{(e)}$ and $\Phi^{*(e)}$
are not affected by the elliptic deformation.

\section{Elliptic DIM algebra as a thermo-field double}
\label{sec:elliptic-dim-algebra}
In this section we prove that elliptic DIM algebra is in fact a
product of two simpler algebras, the conventional (i.e.\
trigonometric) DIM algebra and a Heisenberg algebra. The two parts of
the elliptic DIM algebra are completely decoupled but share a central
charge.

In more physical terms elliptic DIM algebra arises as an equivalent
description of compactified networks of intertwiners of the
trigonometric DIM algebra. These compactified networks correspond to
toric Calabi-Yau three-folds with toric diagrams drawn on a cylinder.
Schematically we can write the relation between the intertwiners of
the two algebras as follows:
\begin{equation}
  \label{eq:60}
  \left[ \Tr ( p^d \Phi(w_1) \Phi(w_2) \cdots \Phi(w_n) )
  \right]_{\mathrm{DIM}} \sim  \left[ \langle \varnothing | \Phi_{\mathrm{ell}}(w_1)
    \Phi_{\mathrm{ell}}(w_2) \cdots \Phi_{\mathrm{ell}}(w_n) | \varnothing \rangle
  \right]_{\text{ell DIM}(p)}
\end{equation}
Here $\Phi(w)$ (resp.\ $\Phi_{\mathrm{ell}}(w)$) are the intertwiners
of DIM (resp.\ elliptic DIM) algebra, $d$ denotes the grading of the
algebra in the direction of the compactification. The parameter $p$ in
the l.h.s. of Eq.~\eqref{eq:60} enters as a box-counting parameter
inside the trace, while in the r.h.s.\ it becomes the parameter of the
elliptic DIM algebra.

Pictorially we can understand the relation~\eqref{eq:60} as a
correspondence between the legs of the toric diagram and the
representations of the elliptic DIM algebra. Recall that before the
compactification a leg of the toric diagram with slope $(n,m)$ used to
correspond to a Fock representation $\mathcal{F}_{q,t^{-1}}^{(n,m)}$
of the trigonometric DIM algebra. $SL(2,\mathbb{Z})$ symmetry acted on
the toric diagram and also on the doublet of indices $(n,m)$ of the
Fock representations.

\begin{figure}[h]
  \centering
  \begin{tabular}{c}
  \includegraphics{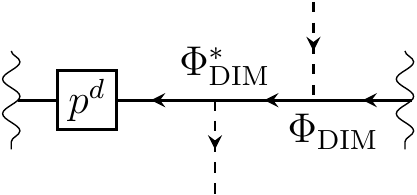} \\
  a) 
\end{tabular}
 \quad
\begin{tabular}{c}
  \includegraphics{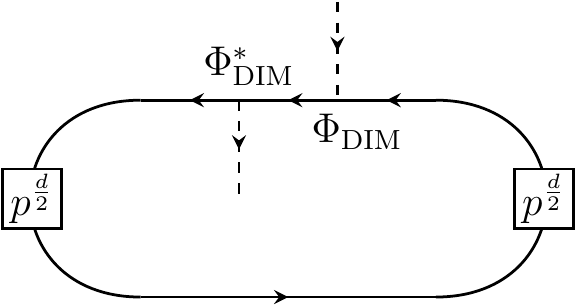}\\
  b) 
\end{tabular}
 \quad
\begin{tabular}{c}
  \includegraphics{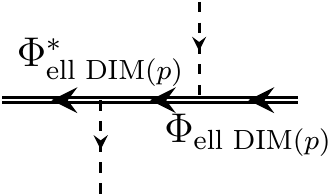}\\
  c)
\end{tabular}
\caption{Three ways of looking at the same compactified network of
  intertwiners. a) Compactified network representing the trace of two
  Higgsed intertwiners $\Phi$ and $\Phi^{*}$ of the
  \emph{trigonometric} DIM algebra. Wavy lines are identified so that
  the network lives on a cylinder. The solid horizontal line denotes
  the Fock representation $\mathcal{F}^{(1,0)}_{q,t^{-1}}$, while the
  vertical dashed lines are vector representations
  $\mathcal{V}_q$. $p$ plays the role of the circumference of the
  compactification circle; accordingly, $p^d$ is the box-counting
  operator. b) The same network as in a) drawn on a plane by
  projecting the cylinder. The box counting operator is split in two
  and the diagram ends with ``cap'' states on the left and right. c)
  The same intertwiner as in a) and b) written in terms of the
  \emph{elliptic} DIM algebra with parameter $p$. In this description
  the ``cap'' states at the ends of the network correspond to vacuum
  states $\langle \varnothing |$ and $| \varnothing \rangle$ in the
  r.h.s.\ of Eq.~\eqref{eq:60}. The horizontal line corresponds to the
  distinguished ``double'' Fock representation, and vertical dashed
  lines are vector representations of the elliptic DIM algebra.}
  \label{fig:1}
\end{figure}

After the compactification we want to keep the rule that a leg in the
diagram corresponds to a representation of the algebra, but now the
algebra is the elliptic DIM algebra. Its representations are
non-isomorphic for different slopes $(n,m)$. Indeed, the slope
corresponding to the direction of the compactification is
distinguished. As we will see this representation is in fact a tensor
product of \emph{two} free boson Fock spaces, whereas the
representation corresponding to the slope orthogonal to the direction
of the compactification is a single Fock space. The reason for this
doubling can be seen in Fig.~\ref{fig:1}\footnote{In our preliminary
  discussion here and also in Fig.~\ref{fig:1} we omit some details
  which don't affect the conceptual picture. In fact the equivalence
  between the intertwiners of the elliptic DIM algebra and those of
  the trigonometric DIM algebra involves a nontrivial Drinfeld twist
  acting on the incoming and outgoing pairs of arrows of each
  intertwiner.}. To reproduce the compactified diagram one needs to
thread the compactified line two times and use special ``cap'' states
on both ends. These states can be understood as the highest weight
states of the horizontal Fock representation of the elliptic DIM
algebra. We will consider the properties of these states in more
detail in sec.~\ref{sec:thermo-field-double}.

We have collected the relevant facts about the elliptic DIM algebra
and its representations in
Appendix~\ref{sec:elliptic-dim-algebra-2}. For a similar reference on
the trigonometric DIM algebra we refer the reader to Appendix A
of~\cite{Zenkevich:2018fzl}.

\subsection{Thermo field double}
\label{sec:thermo-field-double}
In this section we summarize the theory of thermo field double and
show how it can be used to get elliptic deformations of the Heisenberg
algebra. In sec.~\ref{sec:changing-generators} we apply this technique
to the distinguished Heisenberg subalgebra of the \emph{elliptic} DIM
algebra and show that it can naturally be understood as a thermo field
double of a Heisenberg subalgebra of \emph{trigonometric} DIM algebra.

Consider a quantum system with Hilbert space $\mathcal{H}$ and
Hamiltonian $H$. At nonzero temperature $\frac{1}{\beta}$ this system
is described by the thermal density matrix
\begin{equation}
  \label{eq:12}
  \rho_{\mathrm{therm}} = \frac{1}{Z(\beta)} e^{- \beta H},
\end{equation}
where $Z(\beta) = \Tr_{\mathcal{H}} e^{-\beta H}$. Thermal average of
an operator $\mathcal{O}$ acting on $\mathcal{H}$ is computed by
taking a trace with the density matrix:
\begin{equation}
  \label{eq:13}
  \langle \mathcal{O} \rangle = \Tr_{\mathcal{H}}
  ( \rho_{\mathrm{therm}} \mathcal{O} ).
\end{equation}
If the temperature goes to zero, the trace in Eq.~\eqref{eq:13}
becomes the vacuum matrix element:
\begin{equation}
  \label{eq:14}
  \langle \mathcal{O} \rangle \stackrel{\beta \to +\infty}{\to} \langle
  \mathrm{vac} | \mathcal{O} | \mathrm{vac} \rangle.
\end{equation}

One can reformulate the theory in such a way that thermal averages of
operators are also matrix elements (not traces). See e.g.~\cite{Takahasi:1974zn} for an introduction to the technique. The price to
pay is that all the degrees of freedom of the systems are
doubled. Indeed, consider a state
\begin{equation}
  \label{eq:15}
  | \Psi(\beta)\rangle = \frac{1}{\sqrt{Z(\beta)}}\sum_n e^{-\frac{\beta E_n}{2}}|n\rangle
  \otimes |n\rangle \in \mathcal{H} \otimes \mathcal{H},
\end{equation}
where the sum runs over a complete orthonormal\footnote{If the basis
  is orthogonal but not orthonormal one should include the
  normalization factor $\langle n|n\rangle^{-1}$ in each term in the
  sum.} basis of states. We then identify the operator $\mathcal{O}$
with $\mathcal{O} \otimes 1$ and write
\begin{equation}
  \label{eq:16}
  \langle \Psi(\beta) | \mathcal{O} |\Psi(\beta)\rangle =
  \frac{1}{Z(\beta)} \sum_n
  e^{-\beta E_n} \langle n | \mathcal{O} |n\rangle =  \langle
  \mathcal{O} \rangle. 
\end{equation}
Thus, to compute thermal averages one can introduce a fictitious
double of the system, and take matrix element between two judiciously
chosen ``thermal vacuum'' states
$|\Psi(\beta)\rangle$. Fig.~\ref{fig:1} illustrates the reason why the
doubling occurs: the Hilbert space $\mathcal{H}$ is ``threaded two
times'' along the plane to reproduce the trace in~\eqref{eq:13}.

It is remarkable, that for free fields the state $|\Psi(\beta)\rangle$
can in fact be obtained from the vacuum $| \mathrm{vac}\rangle \otimes
| \mathrm{vac} \rangle$ by a Bogolyubov transformation.

\subsubsection{Thermo-field double of a harmonic oscillator}
\label{sec:harmonic-oscillator}
Creation and annihilation operators $\tilde{a}$ and $\tilde{a}^{\dag}$
act on $\mathcal{H}$ in the standard way, so that $[\tilde{a},
\tilde{a}^{\dag}] =1$ and $H = \tilde{a}^{\dag} \tilde{a}$. Then $| \mathrm{vac}
\rangle = |0\rangle$ and an orthonomral basis can be chosen as
\begin{equation}
  \label{eq:17}
  |n\rangle = \frac{(\tilde{a}^{\dag})^n}{\sqrt{n!}} | \varnothing\rangle.
\end{equation} 
We have
\begin{equation}
  \label{eq:19}
  Z(\beta) = \sum_{n \geq 0} e^{-\beta n} = \frac{1}{1 - e^{-\beta}}
\end{equation}
and
\begin{equation}
  \label{eq:18}
  | \Psi(\beta) \rangle = \sqrt{1 - e^{-\beta}} \exp \left( e^{-
      \frac{\beta}{2}} \tilde{a}^{\dag}_{(1)} \tilde{a}^{\dag}_{(2)}  \right)
  |\varnothing\rangle \otimes |\varnothing\rangle, 
\end{equation}
where $\tilde{a}^{\dag}_{(1)} = a^{\dag} \otimes 1$ and
$a^{\dag}_{(2)} = 1 \otimes \tilde{a}^{\dag} $. The unitary Bogolyubov
transformation from the vacuum state $|0 \rangle \otimes |0\rangle$ to
$|\Psi(\beta)\rangle$ is
\begin{equation}
  \label{eq:20}
  U(\beta) = \exp \left( \theta (\tilde{a}^{\dag}_{(1)} \tilde{a}^{\dag}_{(2)} - \tilde{a}_{(1)} \tilde{a}_{(2)})  \right),
\end{equation}
where $\th \theta = e^{- \frac{\beta}{2}}$. We give a short proof of
this statement in Appendix~\ref{sec:thermo-field-double-1}.

One can introduce the operators $a_{(1)}(\beta)$ and $a_{(2)}(\beta)$,
which are Bogolyubov transformations of $a_{(1)} = a \otimes 1$ and
$a_{(2)} = 1 \otimes a$:
\begin{equation}
  \label{eq:21}
  \tilde{a}_{(i)}(\beta) = U(\beta) \tilde{a}_{(i)} U(\beta)^{-1}.
\end{equation}
These ``thermal'' annihilation operators annihilate the thermal
vacuum:
\begin{equation}
  \label{eq:22}
  \tilde{a}_{(i)}(\beta)|\Psi(\beta)\rangle = 0.
\end{equation}
Of course, in a similar way the Bogolyubov transformation can be used
to reexpress the original creation and annihilation operators
$\tilde{a} = \tilde{a}_{(1)}$ and $\tilde{a}^{\dag} =
\tilde{a}_{(1)}^{\dag}$ in terms of the new ``thermal'' operators
$\tilde{a}_{(i)}(\beta)$, $\tilde{a}_{(i)}^{\dag}(\beta)$:
\begin{align}
  \label{eq:31}
  \tilde{a} &= \tilde{a}_{(1)}(\beta) \ch \phi + \tilde{a}_{(2)}^{\dag}(\beta) \sh \phi =
  \frac{1}{\sqrt{1 - e^{-\beta}}} (\tilde{a}_{(1)}(\beta) +
  \tilde{a}_{(2)}^{\dag}(\beta) e^{-\beta/2}),\\
  \tilde{a}^{\dag} &= \tilde{a}_{(1)}^{\dag}(\beta) \ch \phi + \tilde{a}_{(2)}(\beta) \sh \phi=
  \frac{1}{\sqrt{1 - e^{-\beta}}} (\tilde{a}_{(1)}^{\dag}(\beta) +
  \tilde{a}_{(2)}(\beta) e^{-\beta/2}).
\end{align}

It will be convenient for us to rescale the ``thermal'' bosons
introducing
\begin{align}
  \label{eq:32}
  a(\beta) &= \sqrt{1 - e^{-\beta}} \, \tilde{a}_{(1)}(\beta) =
  \tilde{a}_{(1)} +
  \tilde{a}_{(2)}^{\dag} e^{-\beta/2},\\
  b(\beta) &= - e^{\beta/2} \sqrt{1 - e^{-\beta}}\,
  \tilde{a}_{(2)}(\beta) = - e^{\beta/2}\tilde{a}_{(1)}^{\dag} -
  \tilde{a}_{(2)} ,
\end{align}
so that the original generators $\tilde{a}$ are expressed through
$a(\beta)$ and $b(\beta)$ as follows:
\begin{align}
  \label{eq:34}
  \tilde{a} &= \frac{1}{1 - e^{-\beta}} (a(\beta) -
  b^{\dag}(\beta) e^{-\beta}),\\
  \tilde{a}^{\dag} &= \frac{1}{1 - e^{-\beta}} (a^{\dag}(\beta) -
  b(\beta) e^{-\beta}).
\end{align}
The commutation relations for $a(\beta)$, $b(\beta)$
are:
\begin{equation}
  \label{eq:33}
  [a(\beta), a^{\dag}(\beta)] = 1 - e^{-\beta}, \qquad [b(\beta),
  b^{\dag}(\beta)] =  e^{\beta} - 1, \qquad [a(\beta),
  b^{\dag}(\beta)] = [a(\beta),
  b(\beta)] = 0.
\end{equation}
In the low temperature limit $\beta \to +\infty$ the generators
$a(\beta)$, $a^{\dag}(\beta)$ turn into the original generators
$\tilde{a}$, $\tilde{a}^{\dag}$, while $b(\beta)$ and
$b^{\dag}(\beta)$ diverge and need to be rescaled by
$e^{-\beta/2}$. After rescaling they give the decoupled thermal double
generators $a_{(2)}$ and $a_{(2)}^{\dag}$.

\subsubsection{Thermo-field double of a Heisenberg algebra}
\label{sec:thermal-field-double}
Consider the Heisenberg algebra generated by the modes of the free
chiral boson. The algebra is generated by $\tilde{a}_n$, $n \in
\mathbb{Z} \backslash \{ 0 \}$ and the zero modes $Q$ and $P$. The
generators satisfy the commutation relations:
\begin{equation}
  \label{eq:35}
  [\tilde{a}_n, \tilde{a}_m] = n \delta_{n+m,0}, \qquad [Q, P] = 1, \qquad [\tilde{a}_n, P] =
  [\tilde{a}_n, Q] = 0.
\end{equation}
Each of the generators $\tilde{a}_n$ with $n >0$ is the annihilation
operator of a harmonic oscillator, so the thermo-field double in this
case is exactly the same as in sec.~\ref{sec:harmonic-oscillator}. The
conjugate of $\tilde{a}_n$ in this setup is $\tilde{a}_{-n}$. Let us
denote $e^{-\beta} = p$ from now on. We can again express the original
generators $\tilde{a}_n$ in terms of the thermal generators
$a_n(\beta)$ and $b_n(\beta)$ (for brevity we omit the argument
$\beta$ henceforth):
\begin{equation}
  \label{eq:36}
\boxed{  \tilde{a}_n = \frac{1}{1-p^{|n|}} (a_n - b_{-n} p^{|n|})}
\end{equation}
where
\begin{align}
  \label{eq:37}
  [a_n, a_m] &= n (1-p^{|n|}) \delta_{n+m,0}, \\
  [b_n, b_m] &= - n (1-p^{-|n|})
  \delta_{n+m,0},\\
  [a_n, b_m] &= 0. \label{eq:40}
\end{align}

The key aspect of introducing the new ``thermal'' bosons $a_n$ and
$b_n$ is that one can now use the standard Wick's theorem for them to
compute the thermal correlators. Technically, this procedure works
because the normal ordering for $a_n$ and $b_n$ differs form the
normal ordering of the original operators $\tilde{a}_n$. The ``new''
normal ordering produces elliptic functions. Indeed, consider the free
bosonic field
\begin{equation}
  \label{eq:38}
  \tilde{\varphi}(z) = P \ln z + Q - \sum_{n \neq 0} \frac{1}{n} \tilde{a}_n z^{-n}.
\end{equation}
One can build from it the standard primary field $V_{\alpha}(z) =
:e^{\alpha \tilde{\varphi}(z)}:$, where the normal ordering is in
terms of $\tilde{a}_n$. Let us rewrite $V_{\alpha}(z)$ in terms of
$a_n$ and $b_n$ using Eq.~\eqref{eq:36}:
\begin{equation}
  \label{eq:39}
  V_{\alpha}(z) \sim e^{\alpha Q} z^{\alpha P} : \exp \left[ -\alpha \sum_{n \neq 0}
    \frac{z^{-n}}{n(1-p^{|n|})} a_n  \right] \exp \left[
    -\alpha \sum_{n \neq 0}
    \frac{p^{|n|}z^n}{n(1-p^{|n|})} b_n  \right]:,
\end{equation}
where the normal ordering is now in terms of $a_n$, $b_n$ (the
reordering \emph{inside} $V_{\alpha}(z)$ just rescales the vertex
operator by a constant). Using Eqs.~\eqref{eq:37},~\eqref{eq:40} we
get:
\begin{multline}
  \label{eq:41}
  V_{\alpha_1}(z_1) V_{\alpha_2}(z_2) = :V_{\alpha_1}(z_1)
  V_{\alpha_2}(z_2): z_1^{\alpha_1 \alpha_2} \exp \left[ - \alpha_1 \alpha_2\sum_{n \geq 1}
    \frac{1}{n (1-p^n)} \left[ p^n \left( \frac{z_1}{z_2} \right)^n +
      \left(
        \frac{z_2}{z_1} \right)^n  \right]\right] \sim \\
  \sim :V_{\alpha_1}(z_1) V_{\alpha_2}(z_2): z_1^{\alpha_1 \alpha_2}
  \left( \theta_p \left( \frac{z_2}{z_1} \right) \right)^{\alpha_1 \alpha_2}.
\end{multline}
For elliptic DIM algebra we will use a similar rewriting of Heisenberg
generators but in the opposite direction: we will be able to interpret
all the elliptic functions in the elliptic DIM algebra commutation
relations as arising from the ``thermal'' rewriting of a trigonometric
DIM algebra plus an additional ``thermal double'' of its Heisenberg subalgebra.

\subsection{Some hints on the structure of the elliptic DIM algebra}
\label{sec:some-hints-structure}
In this section we give some preliminary considerations which lead us
to the understanding that the elliptic DIM algebra can be understood
in terms of a thermo-field double of a trigonometric DIM algebra. To
this end we look at the unrefined limit $t=q$ of the horizontal Fock
representation $\mathcal{F}^{(1,0)}_{q,t^{-1}}$ of elliptic DIM
algebra. The formulas for this representation are given in
Appendix~\ref{sec:horiz-fock-repr}.

In the unrefined limit the generating functions $\psi^{\pm}(z)$ tend
to the identity and to get the nontrivial generators one needs to
expand them in $(1 - t/q)$ and keep the first order in the
expansion. In this way we obtain:
\begin{align}
  \label{eq:44}
  x^{+}(z) &= e^P :\exp \left[ - \sum_{n \neq 0} \frac{1-q^n}{n(1-p^{|n|})}
    a_n z^{-n} \right] \exp \left[ - \sum_{n \neq 0}
    \frac{1-q^{-n}}{n(1-p^{|n|})}
    b_n p^{|n|} z^n \right]:,\\
  x^{-}(z) &= e^{-P} :\exp \left[ \sum_{n \neq 0} \frac{1-q^n}{n(1-p^{|n|})}
    a_n z^{-n} \right] \exp \left[ \sum_{n \neq 0}
    \frac{1-q^{-n}}{n(1-p^{|n|})}
    b_n p^{|n|} z^n \right]:,\label{eq:68}\\
  \psi^{+}(z) &= 1 - (1 - t/q) \left( \sum_{n \geq 1} \frac{(1 -
      q^n)}{(1-p^n)} a_n z^{-n} + \sum_{n \geq 1} \frac{(1 -
      q^{-n}) p^n}{(1-p^n)}b_n
    z^n\right),\label{eq:70}\\
  \psi^{-}(z) &= 1 + (1 - t/q) \left( \sum_{n \geq 1} \frac{(1 -
      q^{-n})}{(1-p^n)} a_{-n} z^n - \sum_{n \geq 1} \frac{(1 -
      q^n)p^n}{(1-p^n)} b_{-n} z^{-n}\right).\label{eq:71}
\end{align}
One can observe that the modes in the exponents in
$x^{\pm}(z)$~\eqref{eq:44}, \eqref{eq:68} combine precisely
into~\eqref{eq:36}, so that
\begin{align}
  \label{eq:69}
    x^{+}(z) &\sim :\exp \left[ - \sum_{n \neq 0} \frac{1-q^n}{n}
    \tilde{a}_n z^{-n} \right]: = :e^{\tilde{\varphi}(z) - \tilde{\varphi}(z/q)}:,\\
  x^{-}(z) &\sim :\exp \left[ \sum_{n \neq 0} \frac{1-q^n}{n}
    \tilde{a}_n z^{-n} \right]: = :e^{-\tilde{\varphi}(z) + \tilde{\varphi}(z/q)}:,
\end{align}
where $\tilde{\varphi}(z)$ is the ``non-thermal'' free boson
field~\eqref{eq:38}.

The modes of $\psi^{\pm}(z)$~\eqref{eq:70}, \eqref{eq:71}, however,
cannot be rewritten purely as functions of $a_n$. They depend not only
on $a_n$ (i.e.\ $a_n \otimes 1 = a_n^{(1)}$), but also on the
\emph{second} boson $a_n^{(2)}$ in the thermo-field double. This hints
that the ``vertical'' Heisenberg subalgebra inside the elliptic DIM
algebra generated by $\psi^{\pm}(z)$ is the doubled version of that of
trigonometric DIM, whereas all other generators are just rewriting of
the trigonometric expressions.

In the following section we demonstrate that the intuition that we
have obtained for the restricted case of $t=q$ and horizontal Fock
representation is in fact valid more generally. The elliptic DIM
algebra with generic parameters can be explicitly written as a
trigonometric DIM algebra plus a Heisenberg algebra without any
reference to the choice of the representations.

\subsection{Rewriting of the elliptic DIM algebra}
\label{sec:changing-generators}
In this section we show that by rewriting the basis of generators of
the elliptic DIM algebra one can explicitly present it as a direct sum
of a trigonometric DIM algebra and a Heisenberg algebra. Let us remind
that the definitions and notations for elliptic DIM algebra are given
in Appendix~\ref{sec:elliptic-dim-algebra-2}.

\subsubsection{Heisenberg subalgebras}
\label{sec:heis-subalg}
It is evident that (for $\gamma \neq 1$) $H^{\pm}_n$ form two
independent copies of the Heisenberg algebra. Indeed, two sets
$\mathcal{A}_{\pm}= \{ H^{\pm}_n , H^{\mp}_{-n} | n > 0 \}$ commute
between each other and each form a Heisenberg algebra. We would like
to view these Heisenberg algebras as ``thermal'' bosons and perform
the Bogolyubov transform akin to that described in sec.~\ref{sec:thermal-field-double}.

To this end we introduce the new basis of generators:
\begin{equation}
  \label{eq:48}
  \boxed{\begin{aligned}
  \tilde{H}_n &= \sgn(n)(H^{+}_n - H^{-}_n),\\
  \tilde{G}_n &= \sgn(n) (p^{-\frac{n}{2}} H^{+}_{-n} - p^{\frac{n}{2}} H^{-}_{-n}).
\end{aligned}}
\end{equation}
For convenience we also write down the inverse transformation between
the bases:
\begin{align}
  \label{eq:5}
    H^{+}_n &= \frac{\sgn(n)}{1-p^n}(\tilde{H}_n + p^{\frac{n}{2}} \tilde{G}_{-n}),\\
  H^{-}_n &= \frac{\sgn(n) p^n}{1-p^n} (\tilde{H}_n + p^{-\frac{n}{2}} \tilde{G}_{-n}).
\end{align}
The generators $\tilde{H}_n$ and $\tilde{G}_n$ also satisfy two
independent copies of Heisenberg algebras, but now without any
dependence on $p$:
\begin{gather}
  \label{eq:1}
  [\tilde{H}_n, \tilde{H}_m] = [\tilde{G}_n, \tilde{G}_m]  = \delta_{n+m,0} \frac{\kappa_n(\gamma^n
    - \gamma^{-n})}{n},\\
  [\tilde{H}_n, \tilde{G}_m] = 0.
\end{gather}
where $\kappa_n = (1-q^n)(1-t^{-n})(1-t^nq^{-n})$. In what follows we
show that the Heisenberg algebra generated by $\tilde{H}_n$ together
with the \emph{dressed} generators $\tilde{x}^{\pm}(z)$ in fact form a
\emph{trigonometric} DIM algebra, while the second Heisenberg algebra
of $\tilde{G}_n$ decouples completely.

In view of this hypothesis we define the currents of the Heisenberg
subalgebra of the (would be) trigonometric DIM algebra
\begin{equation}
  \label{eq:58}
  \tilde{\psi}^{\pm}(z) = \exp \left[ \sum_{n > 0} \tilde{H}_{\pm n}
    z^{\mp n} \right],
\end{equation}
and the decoupled Heisenberg currents
\begin{equation}
  \label{eq:59}
  \tilde{\phi}^{\pm}(z) = \exp \left[ \sum_{n > 0} \tilde{G}_{\pm n}
    z^{\mp n} \right].
\end{equation}

\subsubsection{Dressing factors}
\label{sec:dressing-factors}
Beyond the vertical Heisenberg subalgebras formed by $H^{\pm}_n$ we
have to modify the generating currents $x^{\pm}(z)$. Indeed, in
elliptic DIM algebra $x^{+}(z)$ satisfy the commutation relations with
elliptic structure functions $G^{\pm}_p(x)$, while the trigonometric
algebra features the trigonometric structure function
\begin{equation}
  \label{eq:2}
  G^{\pm}(x) = G^{\pm}_{p=0}(x) = (1 - q^{\pm 1} x)(1- t^{\mp 1} x)(1
  - t^{\pm 1} q^{\mp 1} x).
\end{equation}

We introduce two dressing factors:
\begin{align}
  \label{eq:6}
  \Gamma_H(z) &= \exp \left[ -\sum_{n \neq 0}
    \frac{p^{|n|}}{(1-p^{|n|})(\gamma^{\frac{n}{2}} +
      \gamma^{-\frac{n}{2}})}\tilde{H}_n  z^{-n} \right],\\
  \Gamma_G(z) &= \exp \left[ -\sum_{n \neq 0}
    \frac{p^{|n|}}{(1-p^{|n|})(\gamma^{\frac{n}{2}} +
      \gamma^{-\frac{n}{2}})} p^{-\frac{|n|}{2}} \tilde{G}_n z^n \right].\label{eq:7}
\end{align}
Using the factors~\eqref{eq:6},~\eqref{eq:7} we define the dressed
currents as follows
\begin{equation}
  \label{eq:3}
  \tilde{x}^{\pm}(z) =  c_{\pm} x^{\pm}(z) \Gamma_H(z) \Gamma_G(z) =  x^{\pm}(z) \exp \left[ -\sum_{n \neq 0}
    \frac{p^{|n|}}{(1-p^{|n|})(\gamma^{\frac{n}{2}} +
      \gamma^{-\frac{n}{2}})}\left(  \tilde{H}_n  +
      p^{-\frac{|n|}{2}} \tilde{G}_{-n}\right)z^{-n} \right],
\end{equation}
where the constants $c_{\pm}$ are chosen so that
\begin{equation}
  \label{eq:62}
  c_{+} c_{-} = \frac{G_p^{-}(1)}{(p;p)_{\infty}^3 G_0^{-}(1)} \exp \left[ \sum_{n\geq
      1} \frac{\kappa_n}{n}\frac{p^n(1-\gamma^n)}{1-p^n} \right],
\end{equation}
where $G_p^{\pm}(x)$ is defined in Eq.~\eqref{eq:9}. The
renormalization constants $c_{\pm}$ depend on the particular way we
order the dressing factors. Their job is to ensure that the
commutation relation~\eqref{eq:4} holds with correct prefactor in
front of the right hand side.

The dressed currents can also be expressed through the old generators
$H^{\pm}_n$:
\begin{equation}
  \label{eq:61}
\boxed{  \tilde{x}^{\pm}(z) = x^{\pm}(z) \exp \left[ - \sum_{n \neq 0}
    \frac{H^{-\sgn(n)}_nz^{-n}}{\gamma^{\frac{n}{2}} + \gamma^{-\frac{n}{2}}}  \right]}
\end{equation}
In this way we can clearly see that in the limit $p\to 0$ the dressing
factors become trivial, since they depend on the generators
$H^{+}_{n<0}$ and $H^{-}_{n>0}$ which vanish in this limit (see
Appendix~\ref{sec:elliptic-dim-algebra-1}). They also trivialize for
$t=q$ because in that case $H_n^{-\sgn(n)} \sim (1-t/q) \to 0$.

The dressing factors $\Gamma_H(z)$, $\Gamma_G(z)$ enjoy the following
commutation relations between themselves and the currents
$\tilde{x}^{\pm}(z)$:
\begin{gather}
  \label{eq:64}
  [\Gamma_H(z), \Gamma_G(w)] = [\tilde{x}^{\pm}(z), \Gamma_G(z)] =
  0,\\
  \Gamma_H(z) \Gamma_H(w) = \exp \left[ \sum_{n \neq 0}
    \frac{\kappa_n}{n} \left( \frac{w}{z} \right)^n
    \frac{p^{2|n|}}{(1-p^{|n|})^2} \frac{\gamma^{\frac{n}{2}} -
      \gamma^{-\frac{n}{2}}}{\gamma^{\frac{n}{2}} +
      \gamma^{-\frac{n}{2}}} \right] \Gamma_H(w) \Gamma_H(z),\\
  \Gamma_G(z) \Gamma_G(w) = \exp \left[ - \sum_{n \neq 0}
    \frac{\kappa_n}{n} \left( \frac{w}{z} \right)^n
    \frac{p^{|n|}}{(1-p^{|n|})^2} \frac{\gamma^{\frac{n}{2}} -
      \gamma^{-\frac{n}{2}}}{\gamma^{\frac{n}{2}} +
      \gamma^{-\frac{n}{2}}} \right] \Gamma_G(w) \Gamma_G(z),\\
  \Gamma_H(z) \tilde{x}^{\pm}(w) = \exp \left[ \pm \sum_{n \neq 0}
    \sgn(n)\frac{\kappa_n}{n} \left( \frac{w}{z} \right)^n
    \frac{p^{|n|} \gamma^{\mp \frac{|n|}{2}}}{1-p^{|n|}} \frac{1}{\gamma^{\frac{n}{2}} +
      \gamma^{-\frac{n}{2}}} \right]
  \tilde{x}^{\pm}(w) \Gamma_H(z).\label{eq:65}
\end{gather}

Using the relations~\eqref{eq:64}--\eqref{eq:65} it is straightforward
to check that the dressed currents obey the ordinary (i.e.\
\emph{trigonometric}) DIM algebra, while the currents
$\tilde{\phi}^{\pm}(z)$ decouple:
\begin{gather}
  \label{eq:66}
  [\tilde{\phi}^{\pm}(z), \tilde{\psi}^{+}(w)] =
  [\tilde{\phi}^{\pm}(z), \tilde{\psi}^{-}(w)] =
  [\tilde{\phi}^{\pm}(z), \tilde{x}^{+}(w)] =
  [\tilde{\phi}^{\pm}(z), \tilde{x}^{-}(w)] =0,  \\
  [\tilde{\psi}^{\pm}(z), \tilde{\psi}^{\pm}(w)] = 0,\qquad
  \tilde{\psi}^{+}(z) \tilde{\psi}^{-}(w) = \frac{g_0\left(\gamma
      \frac{w}{z}\right)}{g_p\left(\gamma^{-1} \frac{w}{z}\right)}
  \tilde{\psi}^{-}(w) \tilde{\psi}^{+}(z),\\
  \tilde{\psi}^{+}(z) \tilde{x}^{\pm}(w) = g_0\left(\gamma^{\mp
      \frac{1}{2}} \frac{w}{z}\right)^{\mp 1} \tilde{x}^{\pm}(w)
  \tilde{\psi}^{+}(z), \qquad \tilde{\psi}^{-}(z) \tilde{x}^{\pm}(w) =
  g_0\left(\gamma^{\mp \frac{1}{2}}
    \frac{w}{z}\right)^{\pm 1} \tilde{x}^{\pm}(w) \tilde{\psi}^{-}(z), \\
  [\tilde{x}^{+}(z), \tilde{x}^{-}(w)] = \frac{1}{G_0^{-}(1)} \left(
    \delta \left( \gamma^{-1} \frac{z}{w} \right) \tilde{\psi}^{+}
    \left( \gamma^{\frac{1}{2}} w \right) - \delta \left( \gamma
      \frac{z}{w} \right) \tilde{\psi}^{-} \left(
      \gamma^{-\frac{1}{2}} w \right) \right), \\
  G_0^{\mp}\left( \frac{z}{w} \right) \tilde{x}^{\pm}(z)
  \tilde{x}^{\pm}(w) = G_0^{\pm}\left( \frac{z}{w} \right)
  \tilde{x}^{\pm}(w) \tilde{x}^{\pm}(z).\label{eq:67}
\end{gather}
Notice that $p$ does not enter the
relations~\eqref{eq:66}--\eqref{eq:67}. In this sense the elliptic
deformation of the DIM algebra is superficial: it can be reproduced by
adding a Heisenberg algebra to the undeformed DIM algebra and then
relabelling the generators.

\subsubsection{Twisting the coproduct and the elliptic \texorpdfstring{$R$}{R}-matrix}
\label{sec:twisting-coproduct-r}
In sec.~\ref{sec:dressing-factors} we have proven the isomorphism of
two algebras: the elliptic DIM algebra and the trigonometric DIM
algebra plus a Heisenberg algebra. However, the algebras on both sides
of the equivalence are in fact Hopf algebras, so it is natural to ask
how their \emph{coproduct} structures are related. Here the situation
is more subtle, because the standard coproduct
structure~\eqref{eq:50}--\eqref{eq:72} on the elliptic DIM algebra
\emph{does not} coincide with the standard coproduct on the
trigonometric DIM algebra. For example, we have
\begin{align}
  \label{eq:73}
  \Delta_{\mathrm{ell}}(H_n^{\pm})&= H^{\pm}_n \otimes \gamma^{\mp
    \frac{n}{2}} + \gamma^{\pm \frac{n}{2}} \otimes H^{\pm}_n,\\
  \Delta_{\mathrm{trig}}(H_n^{\pm})&= H^{\pm}_n \otimes \gamma^{-
    \frac{|n|}{2}} + \gamma^{\frac{|n|}{2}} \otimes
  H^{\pm}_n,\label{eq:75}
\end{align}
where $\Delta_{\mathrm{ell}}$ denotes the
coproduct~\eqref{eq:50}--\eqref{eq:72} on elliptic DIM, and
$\Delta_{\mathrm{trig}}$ is the coproduct obtained from the standard
coproduct on the trigonometric DIM algebra after the change of
variables~\eqref{eq:48}. As one can see the Eq.~\eqref{eq:73} and
Eq.~\eqref{eq:75} are manifestly different.

However, we conjecture that the two coproducts $\Delta_{\mathrm{ell}}$
and $\Delta_{\mathrm{trig}}$ are related by a nontrivial Drinfeld
twist $F$, so that the \emph{coalgebraic} structures on elliptic and
trigonometric DIM algebras are isomorphic. We have not been able to
obtain an explicit expression for $F$, but we think that it can be
written in Khoroshkin-Tolstoy form as an (infinite) product of
elementary twists, the first of which is
\begin{equation}
  \label{eq:76}
  F = \exp \left[ - \sum_{n \geq 1} \frac{n}{\kappa_n} (1-p^{-n})
    \gamma^{\frac{n}{2}} H^{+}_{-n} \otimes \gamma^{-\frac{n}{2}}
    H^{-}_n \right] \times \cdots
\end{equation}
and all other factors lie in the Heisenberg subalgebras with
nonvertical ``slope'' inside the elliptic DIM algebra.

Using the expression~\eqref{eq:76}, incomplete as it is, we can still
find the elliptic $R$-matrix in a tensor product of two vertical
vector representations $\mathcal{V}_q$. To this end we recall that the
$R$-matrix after the Drinfeld twist is given by
\begin{equation}
  \label{eq:77}
  R_{\mathrm{ell}} = F R_{\mathrm{trig}} F_{\mathrm{op}}^{-1},
\end{equation}
where $F_{\mathrm{op}}$ denotes $F$ with the two tensor factors
exchanged. We also notice that all the terms denoted by $\cdots$ in
Eq.~\eqref{eq:76} do not contribute to the $R$-matrix in the vertical
representations, which turns out to be diagonal. Using
Eq.~\eqref{eq:63}, \eqref{eq:78} we then find
\begin{multline}
  \label{eq:74}
  F | _{\mathcal{V}_{q}\otimes \mathcal{V}_q} |f(x_1,x_2)\rangle = \exp \left[ \sum_{n \geq 1} \frac{1}{n} \left(
      \frac{x_{2}}{x_1} \right)^n
    \frac{(1-t^n)(1-(q/t)^n)p^n}{(1-p^n)(1-q^n)} \right]
  |f(x_1,x_2)\rangle =\\
  = \frac{\left( p t
      \frac{x_2}{x_1} ; q,p \right)_{\infty}\left( \frac{pq}{t}
      \frac{x_2}{x_1} ; q,p \right)_{\infty}}{\left( p \frac{x_2}{x_1} ; q,p \right)_{\infty} \left( p q
      \frac{x_2}{x_1} ; q,p \right)_{\infty}}|f(x_1,x_2)\rangle,
\end{multline}
and
\begin{equation}
  \label{eq:82}
  (F_{\mathrm{op}})^{-1} | _{\mathcal{V}_{q}\otimes \mathcal{V}_q} |f(x_1,x_2)\rangle = \frac{\left( p \frac{x_1}{x_2} ; q,p \right)_{\infty} \left( p q
      \frac{x_1}{x_2} ; q,p \right)_{\infty}}{\left( p t
      \frac{x_1}{x_2} ; q,p \right)_{\infty}\left( \frac{pq}{t}
      \frac{x_1}{x_2} ; q,p \right)_{\infty}}|f(x_1,x_2)\rangle,
\end{equation}
where
\begin{equation}
  \label{eq:81}
  (x;q,p)_{\infty} = \prod_{i,j \geq 0} (1 - q^i p^j x).
\end{equation}
Using the expression for the trigonometric $R$-matrix
from~\cite{Zenkevich:2018fzl}
\begin{equation}
  \label{eq:80}
  R_{\mathrm{trig}}| _{\mathcal{V}_{q}\otimes \mathcal{V}_q} |f(x_1,
  x_2)\rangle = \frac{\left(\frac{x_1}{x_2};q\right)_{\infty}\left(q
      \frac{x_1}{x_2};q\right)_{\infty}}{\left(\frac{q x_1}{t
        x_2};q\right)_{\infty}\left(t
      \frac{x_1}{x_2};q\right)_{\infty}}|f(x_1,
  x_2)\rangle
\end{equation}
we get:
\begin{equation}
  \label{eq:79}
  R_{\mathrm{ell}}| _{\mathcal{V}_{q}\otimes \mathcal{V}_q} |f(x_1,
  x_2)\rangle = \frac{\Gamma_{q,p} \left(\frac{q x_1}{t
        x_2}\right) \Gamma_{q,p}\left(t
      \frac{x_1}{x_2}\right)}{\Gamma_{q,p}\left(\frac{x_1}{x_2}\right)
    \Gamma_{q,p}\left(q
      \frac{x_1}{x_2}\right)}|f(x_1,
  x_2)\rangle,
\end{equation}
where
\begin{equation}
  \label{eq:83}
  \Gamma_{q,p}(x) = \frac{\left(\frac{qp}{x};q,p\right)_{\infty}}{(x;q,p)_{\infty}}.
\end{equation}
This coincides with the $R$-matrix obtained from the commutation of
the intertwiners.


\section{Conclusions and discussions}
\label{sec:concl-disc}
In the first part of the paper we have studied the elliptic uplift of
higgsed networks of intertwiners using two different approaches:
either by deforming the DIM algebra and the intertwiners, or by
compactifying the network. We have found the commutation relations of
the elliptically deformed intertwiners and evaluated elliptic
correlators, which reproduce partition functions of $4d$ linear quiver
gauge theories compactified on a two-dimensional torus $T^2$. We have
found that the two approaches give essentially the same partition
functions up to some prefactors associated with flavour
symmetries. These prefactors cancel out when two linear quivers are
glued together and a flavour symmetry subgroup is gauged. We have
found that the elliptic uplifts of the higgsed intertwiners are in
fact the intertwiners of the elliptic deformation of DIM algebra
introduced in~\cite{Saito}.

In the second more abstract part of the paper we have investigated
elliptic DIM algebra itself. Using a dressing transformation of the
generating currents we have found a remarkable isomorphism between
elliptic DIM algebra and a direct sum of two simpler algebras: the
\emph{undeformed} DIM algebra and the Heisenberg algebra. We have also
conjectured that the isomorphism of the algebras extends to an
isomorphism of Hopf algebras and have partly determined the
corresponding Drinfeld twist.

DIM algebra has a $\mathbb{Z}^2$ worth of generators with
$SL(2,\mathbb{Z})$ automorphism group acting on them in a standard
way. Elliptic deformation is not invariant under the
$SL(2,\mathbb{Z})$ transformations, since it depends on the choice of
the compactification direction (or, according to the isomorphism
described in the previous paragraph, on the slope of the ``extra''
Heisenberg algebra added to the DIM algebra). However, DIM algebra
before the deformation already possessed a \emph{preferred direction,}
which is encoded in the choice of the coalgebra structure, i.e.\ the
coproduct. Different choices of the coproduct on DIM lagebra are all
related to each other by $SL(2,\mathbb{Z})$ transformations. However,
after the deformation there appear to be \emph{two} distinguished
directions in the setup: the preferred direction and the direction of
the compactification. Though they both can be rotated by
$SL(2,\mathbb{Z})$, their \emph{relative} slope remains invariant. It
would be interesting to understand if this non-invariance is essential
or can be undone with a cleverly chosen Drinfeld twist. Another
question to ask is what happens if one \emph{doubly} compactifies the
theory. Naively this should restore (or even enhance) the
$SL(2,\mathbb{Z})$ symmetry. However, this remains to be seen.

It is also natural to consider higgsed networks compacitified not
along the horizontal but along the vertical direction. They should
give rise to $3d$ circular quiver gauge theories. One can ask if these
theories are related to the ``horizontally compactified'' $4d$
theories that we have considered here. We can then hypothesize that
there are two ways to deform the DIM algebra: the elliptic one and the
(still missing) \emph{affine} one.  It is possible that there is a
corresponding relation between these two deformations obtained by
compactifying along two orthogonal directions.

\section{Acknowledgements}
\label{sec:concl-disc-1}

We are grateful to A.~Morozov, A.~Mironov and S.~Mironov for
discussions. Y.Z. is partly supported by RFBR grants 19-51-50008-YaF,
19-51-18006-Bolg and 21-52-52004-MNT.


\appendix
\section{Elliptic DIM algebra and some of its representations}
\label{sec:elliptic-dim-algebra-2}
In this appendix we write out the definition of the elliptic DIM
algebra
$U_{q,t,p}^{\mathrm{ell}}(\widehat{\widehat{\mathfrak{gl}}}_1)$, its
coalgebra structure and two types of representations relevant for us
--- the horizontal Fock representation and the vector representation.

\subsection{Elliptic DIM algebra}
\label{sec:elliptic-dim-algebra-1}
The algebra
$U_{q,t,p}^{\mathrm{ell}}(\widehat{\widehat{\mathfrak{gl}}}_1)$ is
generated by the currents $x^{\pm}(z) = \sum_{n \in
  \mathbb{Z}} x^{\pm}_n z^{-n}$, $\psi^{\pm}(z) = \sum_{n \in
  \mathbb{Z}} \psi^{\pm}_n z^{-n}$ and a central element $\gamma$
satisfying the relations\footnote{We rescale the generators with
  respect to~\cite{Saito} to make the symmetry between $q$, $t^{-1}$
  and $t/q$ manifest in the relations: $x^{\pm}_{\mathrm{our}}(z) =
  x^{\pm}_{\cite{Saito}}(z) \theta_p(q^{\mp 1})^{-1} \theta_p(t^{\pm
    1})^{-1}$.}~\cite{Saito}
\begin{gather}
  \label{eq:10}
  [\psi^{\pm}(z), \psi^{\pm}(w)] = 0,\qquad \psi^{+}(z) \psi^{-}(w) =
  \frac{g_p\left(\gamma \frac{w}{z}\right)}{g_p\left(\gamma^{-1}
      \frac{w}{z}\right)}
  \psi^{-}(w) \psi^{+}(z),\\
  \psi^{+}(z) x^{\pm}(w) = g_p\left(\gamma^{\mp \frac{1}{2}}
    \frac{w}{z}\right)^{\mp 1} x^{\pm}(w) \psi^{+}(z), \qquad
  \psi^{-}(z) x^{\pm}(w) = g_p\left(\gamma^{\mp \frac{1}{2}}
    \frac{w}{z}\right)^{\pm 1} x^{\pm}(w) \psi^{-}(z), \label{eq:51}\\
  [x^{+}(z), x^{-}(w)] = \frac{(p;p)_{\infty}^3}{G_p^{-}(1)} \left( \delta \left(
      \gamma^{-1} \frac{z}{w} \right) \psi^{+} \left(
      \gamma^{\frac{1}{2}} w \right) - \delta \left( \gamma
      \frac{z}{w} \right) \psi^{-} \left(
      \gamma^{-\frac{1}{2}} w \right) \right), \label{eq:4}\\
  G_p^{\mp}\left( \frac{z}{w} \right) x^{\pm}(z) x^{\pm}(w) =
  G_p^{\pm}\left( \frac{z}{w} \right) x^{\pm}(w) x^{\pm}(z),\label{eq:24}
\end{gather}
where $\delta(x) = \sum_{n \in \mathbb{Z}} x^n$ and the ``structure
functions'' of the algebra are elliptic deformations of the ordinary
DIM structure functions given by
\begin{gather}
  \label{eq:9}
  G_p^{\pm}(x) = \theta_p(q^{\pm 1}x)\theta_p( t^{\mp 1}x)\theta_p( t^{\pm 1} q^{\mp 1}x), \qquad
  \theta_p(x) = (p;p)_{\infty} (x;p)_{\infty} (p/x;p)_{\infty}, \\
  g(x) = \frac{G_p^{+}(x)}{G_p^{-}(x)}.
\end{gather}
The structure function satisfies $g_p \left( \frac{1}{x} \right) =
\frac{1}{g_p(x)}$ and in particular $g_p(1) = -1$ so that $G_p^{+}(1)
= - G_p^{-}(1)$. Notice that in contrast to the trigonometric case
there are both \emph{positive} and \emph{negative} modes in the
currents $\psi^{\pm}(z)$. Just as the trigonometric DIM algebra, the
elliptic one is manifestly symmetric under any permutation of the
parameters $q$, $t^{-1}$, and $\frac{t}{q}$.

In the limit $p\to 0$ elliptic structure function turns into
trigonometric one:
\begin{equation}
  \label{eq:11}
  G^{\pm}(x) \stackrel{p \to 0}{\to} (1 - q^{\pm 1}x)(1 - t^{\mp
    1}x)(1 - t^{\pm 1} q^{\mp 1}x),
\end{equation}
so that the conventional DIM algebra is recovered.

It is convenient to introduce the generators $H^{\pm}_n$ such
that\footnote{Both sides should be understood as formal power series
  in $p$. In this way only a finite number of terms from the exponent
  contribute to each $\psi^{\pm}_n$ at a given order in $p$.}
\begin{equation}
  \label{eq:46}
  \psi^{\pm}(z) = \psi^{\pm}_0 \exp \left[ \sum_{n \neq 0} H^{\pm}_n z^{-n} \right].
\end{equation}
The elements $\psi^{\pm}_0$ are central. Notice that, unlike the
trigonometric case the index $n \neq 0$ can be both positive and
negative for $H^{\pm}_n$. We can rewrite the commutation
relations~\eqref{eq:10}--\eqref{eq:51} in terms of $H_n^{\pm}$:
\begin{gather}
  \label{eq:47}
  [H^{\pm}_n, H^{\pm}_m] = 0, \qquad \qquad [H^{+}_n , H^{-}_m] =
  \delta_{n+m,0} \frac{\kappa_n}{n(1-p^n)} (\gamma^n - \gamma^{-n}),\\
  [H^{+}_n, x^{\pm}(z)] = \mp \frac{\kappa_n}{n (1-p^n)}
  \gamma^{\mp \frac{n}{2}} z^n x^{\pm}(z),\\
  [H^{-}_n, x^{\pm}(z)] = \mp \frac{\kappa_n p^n}{n (1-p^n)}
  \gamma^{\pm \frac{n}{2}} z^n x^{\pm}(z). \label{eq:57}
\end{gather}
where $\kappa_n = (1-q^n)(1-t^{-n})(1-t^nq^{-n}) = - \kappa_{-n}$. In
the limit $p \to 0$ the fate of the generators $H^{\pm}_n$ can be read
from the commutation relations~\eqref{eq:47} and depends on the sign
of the index $n$:
\begin{enumerate}
\item $H^{+}_{n > 0}$ and $H^{-}_{n<0}$ remain finite and become the
modes of the ``vertical'' Heisenberg subalgebra of the trigonometric
DIM algebra.

\item $H^{+}_{n < 0}$ and $H^{-}_{n>0}$ vanish as $\sqrt{p}$ in the
$p \to 0$ limit. Naively, they disappear completely so that only
\emph{half} of the ``elliptic'' modes $H^{\pm}_n$ remain in the
trigonometric limit. However, if one rescales the vanishing modes by
$p^{-1/2}$ they survive and form an extra Heisenberg algebra. See
sec.~\ref{sec:heis-subalg} for the complete description of this
Heisenberg subalgebra valid for general $p$.
\end{enumerate}

In writing Eqs.~\eqref{eq:47}--\eqref{eq:57} one has to be careful,
since there are two ways to write the structure function $g_p(x)$ as
an exponential (they differ by an infinite sum of delta-functions):
  \begin{equation}
    \label{eq:8}
    g_p(x) = \exp \left[ \sum_{n \neq  0} \frac{\kappa_n}{n} \frac{x^n}{1-p^n} \right],
  \end{equation}
  or
  \begin{equation}
    \label{eq:56}
    (g_p(x^{-1}))^{-1} = \exp \left[ \sum_{n \neq  0}
      \frac{\kappa_n}{n} \frac{p^n x^n}{1-p^n} \right].
  \end{equation}
  To determine which one is needed in each commutation relation one
  can check the limit $p \to 0$. According to Eq.~\eqref{eq:47}

There exists a coproduct on the elliptic DIM algebra given by:
\begin{align}
  \label{eq:50}
  \Delta(\gamma) &= \gamma \otimes \gamma,\\
  \Delta(\psi^{\pm}(z)) &= \psi^{\pm}(\gamma_{(2)}^{\pm \frac{1}{2}} z)
  \otimes \psi^{\pm}(\gamma_{(1)}^{\mp \frac{1}{2}} z),\label{eq:52}\\
  \Delta(x^{+}(z)) &= x^{+}(z) \otimes 1 +
  \psi^{-}(\gamma_{(1)}^{\frac{1}{2}} z) \otimes x^{+}(\gamma_{(1)} z),\\
  \Delta(x^{-}(z)) &= x^{-}(\gamma_{(2)}z) \otimes
  \psi^{+}(\gamma_{(2)}^{\frac{1}{2}} z) + 1 \otimes x^{-}(z),\label{eq:72}
\end{align}
where $\gamma_{(1)} = \gamma \otimes 1$ and $\gamma_{(2)} = 1 \otimes
\gamma$. In terms of the generators $H_n^{\pm}$ the
coproduct~\eqref{eq:52} reads
\begin{equation}
  \label{eq:53}
  \Delta(H_n^{\pm}) = H^{\pm}_n \otimes \gamma^{\mp \frac{n}{2}} +
  \gamma^{\pm \frac{n}{2}} \otimes H_n^{\pm}.
\end{equation}

For completeness let us also give the expressions for the counit
$\epsilon$
\begin{gather}
  \label{eq:54}
  \epsilon(x^{\pm}(z)) = \epsilon(H_n^{\pm}) = 0,\\
  \epsilon(\gamma) = \epsilon(\psi^{\pm}_0) = 1.
\end{gather}
and (formal) antipode $S$ of the elliptic DIM algebra:
\begin{align}
  \label{eq:55}
  S(\gamma) &= \gamma^{-1},\\
  S(\psi^{\pm}(z)) &= (\psi^{\pm}(z))^{-1},\\
  S(x^{+}(z)) &= -\left(\psi^{-}(\gamma^{- \frac{1}{2}}z)\right)^{-1}
  x^{+}(\gamma^{-1}z),\\
  S(x^{-}(z)) &= - x^{-}(\gamma^{-1}z) \left(\psi^{+}(\gamma^{- \frac{1}{2}}z)\right)^{-1}.
\end{align}

\subsection{Vector representation}
\label{sec:vect-repr}
Vector representation $\mathcal{V}_q$ is the representation of the
elliptic DIM with trivial central charges on the space of functions~\cite{Wang}.
\begin{align}
  \label{eq:42}
  x^{+}(z) f(x) &= -\frac{1}{\theta_p(q^{-1})} \delta \left(
    \frac{x}{qz}
  \right) q^{-x \partial_x} f(x), \\
  x^{-}(z) f(x) &= -\frac{1}{\theta_p(q)} \delta \left( \frac{x}{z}
  \right) q^{x \partial_x} f(x), \\
  \psi^{+}(z) f(x) &= \frac{\theta_p \left( \frac{tx}{qz} \right)
    \theta_p \left( \frac{x}{tz} \right)}{\theta_p \left( \frac{x}{z}
    \right) \theta_p \left( \frac{x}{q z} \right)}
  f(x),\\
  \psi^{-}(z) f(x) &= \frac{\theta_p \left( \frac{qz}{tx} \right)
    \theta_p \left( \frac{t z}{x} \right)}{\theta_p \left( \frac{z}{x}
    \right) \theta_p \left( q \frac{z}{x} \right)} f(x),\\
  \gamma f(x) &= f(x).
\end{align}
There are three vector representations obtained by exchanging $q \to
t^{-1} \to \frac{t}{q}$. In terms of $H^{\pm}_n$ generators we have
\begin{align}
  \label{eq:63}
  H_n^{+}f(x) &= \frac{x^n (1-t^{-n}) \left( 1 - \left( \frac{t}{q}
      \right)^n \right)}{n(1-p^n)} f(x),\\
  H_n^{-}f(x) &= \frac{x^n p^n(1-t^{-n}) \left( 1 - \left( \frac{t}{q}
      \right)^n \right)}{n(1-p^n)} f(x).\label{eq:78}
\end{align}

\subsection{Horizontal Fock representation}
\label{sec:horiz-fock-repr}
In this representation $\mathcal{F}_{t^{-1},t/q}$~\cite{Saito} DIM
algebra acts on the tensor product of two free boson Fock spaces
generated by the modes $a_n$ and $b_n$ satisfying the commutation
relations
\begin{align}
  \label{eq:45}
  [a_n, a_m] &= n \delta_{n+m,0} (1-p^{|n|})\frac{1 -
    q^{|n|}}{1 - t^{|n|}},\\
  [b_n, b_m] &= - n \delta_{n+m,0} (1-p^{-|n|}) \frac{1 -
    q^{-|n|}}{1 - t^{-|n|}},\\
  [a_n, b_m] &= 0.
\end{align}
We also introduce the zero modes $Q$ and $P$ commuting with $a_n$ and
$b_n$ and satisfying
\begin{equation}
  [P,Q]=1.
\end{equation}
$P$ acts diagonally on $\mathcal{F}_{t^{-1},t/q}$ and gives the
spectral parameters of the Fock representation:
\begin{equation}
  e^P |_{\mathcal{F}_{t^{-1},t/q}(u)} = u.
\end{equation}
The operator $Q$ acts between \emph{different} Fock representations by
shifting the spectral parameter. We use it in the intertwiner 

The elliptic DIM algebra generators act as follows:
\begin{align}
  \label{eq:43}
  x^{+}(z) &= \frac{(p;p)_{\infty}^3 e^P}{\theta_p(q^{-1}) \theta_p(t)} :\exp \left[ - \sum_{n \neq 0} \frac{1-t^n}{n(1-p^{|n|})}
  a_n z^{-n} \right] \exp \left[ - \sum_{n \neq 0} \frac{1-t^{-n}}{n(1-p^{|n|})}
  b_n p^{|n|} z^n \right]:,\\
  x^{-}(z) &= \frac{(p;p)_{\infty}^3 e^{-P}}{\theta_p(q^{-1}) \theta_p(t^{-1})} :\exp \left[ \sum_{n \neq 0} \left( \frac{t}{q} \right)^{\frac{|n|}{2}} \frac{1-t^n}{n(1-p^{|n|})}
  a_n  z^{-n} \right] \exp \left[ \sum_{n \neq 0} \left( \frac{t}{q} \right)^{-\frac{|n|}{2}} \frac{1-t^{-n}}{n(1-p^{|n|})}
  b_n p^{|n|} z^n \right]:,\\
\psi^{+}(z) &= \exp \left[ - \sum_{n \geq 1} \frac{(1 - t^n)(1 - t^n/q^n)}{n(1-p^n)}
   \left( \frac{q}{t} \right)^{\frac{n}{4}} a_n
z^{-n}\right] \exp \left[ - \sum_{n \geq 1} \frac{(1 - t^{-n})(1 -
  q^n/t^n) p^n}{n(1-p^n)}
   \left( \frac{t}{q} \right)^{\frac{n}{4}} b_n
z^n\right],\\
\psi^{-}(z) &= \exp \left[ \sum_{n \geq 1} \frac{(1 - t^{-n})(1 - t^n/q^n)}{n(1-p^n)} \left( \frac{q}{t} \right)^{\frac{n}{4}} a_{-n}
z^n\right] \exp \left[ \sum_{n \geq 1} \frac{(1 - t^n)(1 - q^n/t^n)p^n}{n(1-p^n)}
    \left( \frac{t}{q} \right)^{\frac{n}{4}} b_{-n}
z^{-n}\right],\\
\gamma &= \sqrt{\frac{t}{q}},
\end{align}
or, in terms of $H^{\pm}_n$ generators:
\begin{align}
  \label{eq:49}
  H^{+}_n &= - \frac{(1-t^n) \left( 1 - \left( \frac{t}{q} \right)^n
    \right)}{n (1 - p^n)} \left( \frac{q}{t} \right)^{\frac{n}{4}}
  \begin{cases}
    a_n, & n>0,\\
    b_{-n}, & n<0,
  \end{cases}\\
  H^{-}_n &= - \frac{(1-t^n) \left( 1 - \left( \frac{q}{t} \right)^n
    \right)}{n (1 - p^{-n})} \left( \frac{t}{q} \right)^{\frac{n}{4}}
  \begin{cases}
    b_{-n}, & n>0,\\
    a_n, & n<0.
  \end{cases}
\end{align}

\section{Thermo-field double as a Bogolyubov transformation}
\label{sec:thermo-field-double-1}
Let us prove that the thermo-field vacuum state~\eqref{eq:18} can be
obtained from the standard vacuum $|0\rangle \otimes |0\rangle$ by the
Bogolyubov transformation~\eqref{eq:20}. Indeed, we notice that
\begin{equation}
  \label{eq:23}
  (a^{\dag}_{(1)} a^{\dag}_{(2)} - a_{(1)} a_{(2)}) e^{\psi
    a^{\dag}_{(1)} a^{\dag}_{(2)} }
  |\varnothing\rangle \otimes |\varnothing\rangle = \left( (1-\psi^2)\frac{\partial}{\partial \psi}
    - \psi   \right) e^{\psi
    a^{\dag}_{(1)} a^{\dag}_{(2)} }
  |\varnothing\rangle \otimes |\varnothing\rangle.
\end{equation}
We introduce the new coordinate $\phi (\psi)$ such that
\begin{equation}
  \label{eq:25}
  (1-\psi^2)\frac{\partial}{\partial \psi} = \frac{\partial}{\partial \phi},
\end{equation}
and therefore
\begin{equation}
  \label{eq:26}
  \psi(\phi) = \th \phi.
\end{equation}
We have
\begin{multline}
  \label{eq:27}
  (a^{\dag}_{(1)} a^{\dag}_{(2)} - a_{(1)} a_{(2)}) e^{\psi
    a^{\dag}_{(1)} a^{\dag}_{(2)} }
  |\varnothing\rangle \otimes |\varnothing\rangle = \left( \frac{\partial}{\partial \phi}
    - \th \phi   \right) e^{\th (\phi)
    a^{\dag}_{(1)} a^{\dag}_{(2)} }
  |\varnothing\rangle \otimes |\varnothing\rangle =\\
  = \ch (\phi) \frac{\partial}{\partial \phi} \left[ \frac{1}{\ch \phi} e^{\th (\phi)
    a^{\dag}_{(1)} a^{\dag}_{(2)} }
  |0\rangle \otimes |0\rangle \right],
\end{multline}
So that
\begin{multline}
  \label{eq:28}
  \exp (\theta(a^{\dag}_{(1)} a^{\dag}_{(2)} - a_{(1)} a_{(2)})) e^{\psi
    a^{\dag}_{(1)} a^{\dag}_{(2)} }
  |\varnothing\rangle \otimes |\varnothing\rangle = \ch \phi \exp \left(
    \theta \frac{\partial}{\partial \phi} \right) \left[ \frac{1}{\ch \phi} e^{\th (\phi)
    a^{\dag}_{(1)} a^{\dag}_{(2)} }
  |0\rangle \otimes |0\rangle \right] =\\
= \frac{\ch \phi}{\ch (\phi +
  \theta)} e^{\th (\phi + \theta)
    a^{\dag}_{(1)} a^{\dag}_{(2)} }
  |\varnothing\rangle \otimes |\varnothing\rangle.
\end{multline}
Setting $\psi = \phi =0$ in Eq.~\eqref{eq:28} we get
\begin{equation}
  \label{eq:29}
 \exp (\theta (a^{\dag}_{(1)} a^{\dag}_{(2)} - a_{(1)} a_{(2)})) 
  |\varnothing\rangle \otimes |\varnothing\rangle = \frac{1}{\ch 
  \theta} e^{\th (\theta)
    a^{\dag}_{(1)} a^{\dag}_{(2)} }
  |0\rangle \otimes |0\rangle.
\end{equation}
Combining Eq.~\eqref{eq:29} and Eq.~\eqref{eq:18} we can write
\begin{equation}
  \label{eq:30}
  U(\beta) |\varnothing \rangle \otimes |\varnothing \rangle = |\Psi(\beta)\rangle.
\end{equation}

\section{Appendix}
\label{sec:appendix}

In this Appendix, we would like to collect some definitions and relations we used in the main text as well as to present some technical details omitted in the main text for clarification. Two elliptic functions repeatedly appeared in our computations, one of them is the elliptic gamma function $\Gamma_{p,q}(z)$ (also sometimes denoted as $\Gamma(z)$ to avoid notational cluttering) defined as

\begin{align}
\Gamma_{p,q}(z)\equiv \prod_{i,j=0}^{\infty}\frac{1-z^{-1}\,q^{i+1}p^{j+1}}{1-z\,q^{i}p^{j}}=\frac{(qpz^{-1};q,p)_{\infty}}{(z;q,p)_{\infty}},
\end{align}
where we have used the double $pq$-Pochhammer symbol whose definition is clear in the above expression. The other one is the (reduced) elliptic Jacobi $\theta$-function $\theta_q(z)$,

\begin{align}
\theta_q(z)=(q;q)_{\infty}(z;q)_{\infty}(qz^{-1};q)_{\infty},
\end{align}
with the usual $q$-Pochhammer symbol. The elliptic gamma function enjoys the so-called reflection property,
\begin{align}
\Gamma_{p,q}(z)=\frac{1}{\Gamma_{p,q}(qp z^{-1})}.
\end{align}

Two other relations between elliptic gamma function and theta function prove to be extremely useful to match the expressions we obtain from DIM algebraic evaluations to localization computations in gauge theory:
\begin{align}
\Gamma_{p,q}(qz)=\frac{\theta_p(z)}{(p;p)_{\infty}}\Gamma_{p,q}(z),\qquad \Gamma_{p,q}(pz)=\frac{\theta_q(z)}{(q;q)_{\infty}}\Gamma_{p,q}(z).
\end{align}
We also make use of the following $\theta$-function identities ,

\begin{align}
\theta_q(z)=(-z)\,\theta_q(z^{-1}),\qquad \theta_q(qz)=(-z)^{-1}\,\theta_{q}(z).
\end{align}

\end{document}